\documentclass[fleqn,usenatbib,useAMS,onecolumn]{mnras}

\usepackage{graphicx}	
\usepackage{amsmath}	
\usepackage{amssymb}	
\usepackage{multicol}        
\usepackage{bm}		
\usepackage{pdflscape}	
\usepackage[T1]{fontenc}
\usepackage{ae,aecompl}
\usepackage{newtxtext,newtxmath}

\title[SN 2018hti: a nearby superluminous supernova]{SN 2018hti: a nearby superluminous supernova discovered in a metal-poor galaxy}

\author[W. L. Lin et al.]{W. L. Lin$^{1}$, X. F. Wang$^{1,2}$, W. X. Li$^{1}$, J. J. Zhang$^{3,4,5}$, J. Mo$^{1}$, H. N. Sai$^{1}$, X. H. Zhang$^{1}$, \newauthor   
A. V. Filippenko$^{6,7}$, W. K. Zheng$^{6}$, T. G. Brink$^{6}$, E. Baron$^{8}$, J. M. DerKacy$^{8}$, \newauthor 
S. A. Ehgamberdiev$^{9}$, D. Mirzaqulov$^{9}$, X. Li$^{1}$, J. C. Zhang$^{1}$, S. Y. Yan$^{1}$, G. B. Xi$^{1}$, \newauthor 
Y. Hsiao$^{1}$, T. M. Zhang$^{10,11}$, L. J. Wang$^{12}$, L. D. Liu$^{13}$, D. F. Xiang$^{1}$, C. Y. Wu$^{1}$, \newauthor 
L. M. Rui$^{1}$, Z. H. Chen$^{1}$
\\
$^{1}$Physics Department and Tsinghua Center for Astrophysics (THCA), Tsinghua University, Beijing 100084, China; linwl@mail.tsinghua.edu.cn; \\
wang\_xf@mail.tsinghua.edu.cn\\
$^{2}$Beijing Planetarium, Beijing Academy of Science and Technology, Beijing 100044, China\\
$^{3}$Yunnan Observatories, Chinese Academy of Sciences, Kunming 650216, China\\
$^{4}$Key Laboratory for the Structure and Evolution of Celestial Objects, Chinese Academy of Sciences, Kunming 650216, China\\
$^{5}$Center for Astronomical Mega-Science, Chinese Academy of Sciences, 20A Datun Road, Chaoyang District, Beijing 100012, China\\
$^{6}$Department of Astronomy, University of California, Berkeley, CA 94720-3411, USA\\
$^{7}$Miller Senior Fellow, Miller Institute for Basic Research in Science, University of California, Berkeley, CA 94720, USA\\
$^{8}$Homer L. Dodge Department of Physics and Astronomy, University of Oklahoma, Norman, OK 73019, USA\\
$^{9}$Ulugh Beg Astronomical Institute, Uzbekistan Academy of Sciences, Tashkent 100052, Uzbekistan\\
$^{10}$Key Laboratory of Optical Astronomy, National Astronomical Observatories, Chinese Academy of Sciences, Beijing 10101, China\\
$^{11}$School of Astronomy and Space Science University of Chinese Academy of Sciences,  Beijing 101408, China\\
$^{12}$Astroparticle Physics, Institute of High Energy Physics, Chinese Academy of Sciences, Beijing 100049, China\\
$^{13}$Department of Astronomy, Beijing Normal University, Beijing 100875, China}

\begin{document}
\label{firstpage}
\pagerange{\pageref{firstpage}--\pageref{lastpage}}
\maketitle
\begin{abstract}
SN 2018hti is a Type I superluminous supernova (SLSN~I) with an absolute $g$-band magnitude of $-22.2$ at maximum brightness, discovered by ATLAS in a metal-poor galaxy at a redshift of 0.0612. We present extensive photometric and spectroscopic observations of this supernova, covering the phases from $\sim -35$ d to more than +340 d from the $r$-band maximum. Combining our $BVgri$-band photometry with  {\it Swift} UVOT optical/ultraviolet photometry, we calculated the peak luminosity as $\sim 3.5\times10^{44}$ erg s$^{-1}$. Modeling the observed light curve reveals that the luminosity evolution of SN 2018hti can be produced by an ejecta mass of 5.8 M$_\odot$ and a magnetar with a magnetic field of $B=1.8\times10^{13}$~G having an initial spin period of $P_0=1.8$ ms. Based on such a magnetar-powered scenario and a larger sample, a correlation between the spin of the magnetar and the kinetic energy of the ejecta can be inferred for most SLSNe~I, suggesting a self-consistent scenario. Like for other SLSNe~I, the host galaxy of SN 2018hti is found to be relatively faint ($M_{g} = -17.75$ mag) and of low metallicity ($Z=0.3~Z_\odot$), with a star formation rate of 0.3 M$_\odot$ yr$^{-1}$. According to simulation results of single-star evolution, SN 2018hti could originate from a massive, metal-poor star with a zero-age main sequence (ZAMS) mass of 25--40 M$_\odot$, or from a less massive rotating star with $M_\mathrm{ZAMS} \approx 16$--25 M$_\odot$. For the case of a binary system, its progenitor could also be a star with $M_\mathrm{ZAMS} \gtrsim 25$ M$_\odot$.

\end{abstract}


\section{Introduction}

Superluminous supernovae (SLSNe) represent a rare subtype of supernovae (SNe) that are much brighter than ordinary stellar explosions (e.g., \citealp{2012Sci...337..927G, 2019ARA&A..57..305G, 2019NatAs...3..697I}). Type~I SLSNe (SLSNe~I) lack prominent hydrogen features around peak brightness, as distinguished from Type II SLSNe (SLSNe~II). Before maximum light, the spectra of SLSNe~I are characterized by blue continua and lines of ionized oxygen and carbon (e.g., \citealp{2011Natur.474..487Q, 2016MNRAS.458.3455M, 2016ApJ...826...39N, 2018ApJ...855....2Q, 2019ApJ...882..102G}). The five broad O\,\textsc{ii} absorption complexes at 3500--5000\,\AA, especially the W-shaped features, are typical of the early-time spectra of SLSNe~I. After peak brightness, SLSNe~I eventually evolve to be spectrally similar to normal and broad-lined SNe~Ic (e.g., \citealp{2010ApJ...724L..16P, 2016ApJ...828L..18N, 2017ApJ...845...85L, 2019ApJ...872...90B}). 

Detailed studies of SLSNe~I (e.g., \citealp{2018ApJ...860..100D, 2018ApJ...855....2Q}) have shown that a spectroscopic similarity appears to be shared by those with peak absolute magnitude brighter than $M_\mathrm{g}=-19.8$ mag, which is then suggested as a photometric threshold for SLSNe~I \citep{2019ARA&A..57..305G}. Such extraordinary luminosity can be explained by a magnetar-powered scenario where a rapidly rotating newly-born magnetar dissipates its rotational energy via magnetic dipole radiation to power the SN (e.g., \citealp{2007ApJ...666.1069M, 2010ApJ...717..245K, 2010ApJ...719L.204W, 2016ApJ...821...22W}). This model provides good fits to the diverse light curves of most SLSNe~I (e.g., \citealp{2013ApJ...770..128I, 2014MNRAS.444.2096N, 2015MNRAS.452.3869N, 2015ApJ...799..107W, 2017ApJ...842...26L, 2017ApJ...850...55N, 2017ApJ...840...12Y}). In addition, it can explain the radio emission spatially coincident with SN 2010md (also known as PTF10hgi), which could be produced by a wind nebula around a young magnetar \citep{2019ApJ...876L..10E}. The bright X-ray emission associated with rare SLSNe~I (SCP60F6 and PTF12dam) also implies the possible presence of a central engine \citep{2013ApJ...771..136L, 2018ApJ...864...45M}.

Some SLSNe~I exhibit a sharp bolometric peak or post-peak undulations, which invoke the interaction of ejecta with circumstellar material (CSM;  e.g., \citealp{2012ApJ...746..121C, 2013ApJ...773...76C, 2017ApJ...851L..14W, 2018ApJ...856...59L}). The observations of strong H$\alpha$ emission in late-time spectra of a few SLSNe~I indicate interaction between ejecta and a hydrogen-rich shell far from the site of the explosion, at $\sim10^{16}$ cm \citep{2015ApJ...814..108Y, 2017ApJ...848....6Y}. The fallback accretion onto a black hole (BH) can serve as an alternative channel to power SLSNe (e.g., \citealp{2013ApJ...772...30D, 2018ApJ...867..113M, 2018MNRAS.475L..11M}).

SN 2018hti is a good example for examining the central engine and tracing the progenitor properties of SLSNe~I, since it is one of the closest slowly evolving SLSNe discovered to date and hence allowed follow-up observations for more than 1~yr. In this paper, we present observations and modeling analysis of this SLSN~I. The observations of SN 2018hti are presented in Section \ref{Sec: Obs}. We provide details of its spectral evolution, spanning from extremely early times to the nebular phase, and discuss its spectral similarity to other SLSNe~I and well-observed SNe~Ic. In Section \ref{Sec: Photometry}, we perform a detailed analysis of the photometric evolution based on the magnetar-powered model. We summarize in Section \ref{Sec: Concl}. Throughout this paper, the the lambda cold dark matter ($\Lambda$CDM) cosmological model with $\Omega_{M} = 0.27$, $\Omega_{\Lambda} = 0.73$, and H$_{0}$ = 73 km s$^{-1}$ Mpc$^{-1}$ is adopted when calculating the distance modulus.

\section{Observations}
\label{Sec: Obs}
\subsection{Photometry}

\begin{figure}
\center
\includegraphics[angle=0,width=1\textwidth]{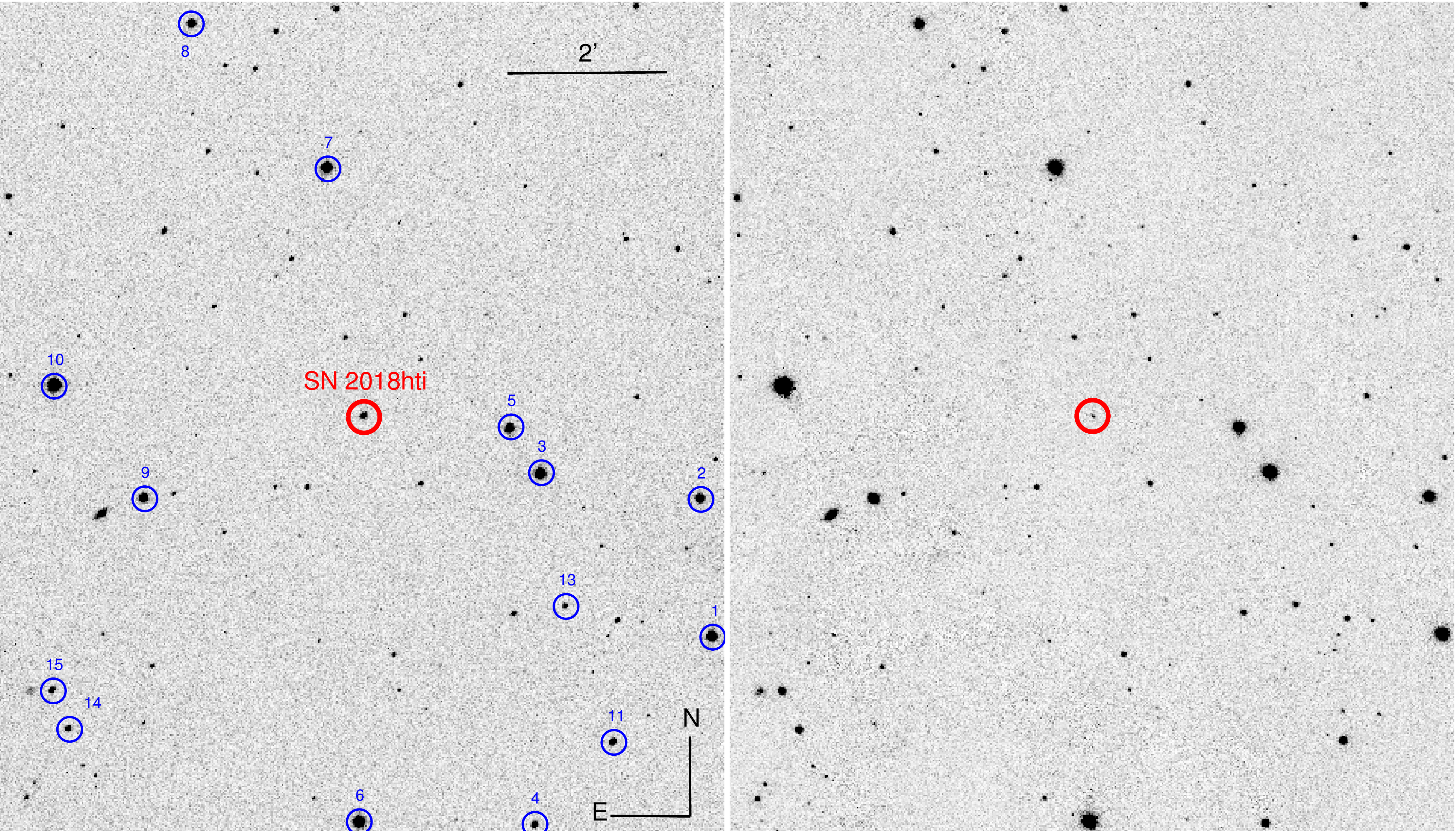}
\caption{Left panel: the $r$-band image of SN 2018hti, observed on 2019 January 2 with the Tsinghua-NAOC 80\,cm telescope. This image shows the position of SN 2018hti (red circle) as well as most of the local standard stars (blue circles; Table~\ref{Table: referstar}) used to calibrate the magnitudes. Right panel: the $r$-band pre-explosion image from Pan-STARRS. Comparison of the two images illustrates that the SN is coincident with a dwarf galaxy. North is up and east to the left.}
\label{fig: FinderChart}
\end{figure}

SN 2018hti was discovered on 2018 November 02.52 (UT dates are used throughout  this paper; JD = 2458425.018) at $\alpha = 03^{\rm h}40^{\rm m}53.754^{\rm s}$ and $\delta = +11\degr 46\arcmin 37.29''$ (J2000) by the Asteroid Terrestrial-impact Last Alert System (ATLAS; \citealp{2018PASP..130f4505T, 2020PASP..132h5002S}), with an AB magnitude of 18.43 in the orange ($o$) band \citep{2018TNSTR1680....1T}. The finder chart showing the SN and most of the local standard stars (Table~\ref{Table: referstar}) is given in Figure~\ref{fig: FinderChart}. A spectrum taken a few days after discovery indicates that SN 2018hti was an SLSN~I that is a good match with PTF09cnd \citep{2018ATel12183....1A}. A redshift of $z=0.062$ is measured from the narrow host-galaxy emission lines of H and [O\,\textsc{iii}] \citep{2018ATel12183....1A}. The Galactic reddening inferred for SN 2018hti is found to be $E(B-V)=0.4$ mag from the NED \citep{2011ApJ...737..103S}, corresponding to a $V$-band extinction of 1.3 mag with the \citet{1989ApJ...345..245C} extinction law ($R_{V} = 3.1$).

Our follow-up photometric observations were collected with the Lijiang 2.4-m telescope (LJT; \citealp{2015RAA....15..918F}) at Yunnan Astronomical observatory and the Tsinghua-NAOC 80-cm telescope (TNT; \citealp{2008ApJ...675..626W, 2012RAA....12.1585H}) at NAOC Xinglong Observatory. Both the LJT and TNT observations were obtained in the Johnson-Cousins $BV$ and Sloan $gri$ bands. Standard \textsc{iraf}\footnote{\textsc{iraf} is distributed by the National Optical Astronomy Observatories, which are operated by the Association of Universities for Research in Astronomy, Inc., under cooperative agreement with the U.S. National Science Foundation (NSF).} routines were adopted to preprocess the CCD images of LJT and TNT, including corrections for the bias and flat field, and removal of cosmic rays. The colour terms of the TNT and LJT were taken from \citet{2012RAA....12.1585H} and the extinction coefficients at the corresponding sites were measured by observing \citet{1992AJ....104..340L} standards during the nights \citep{2016ApJ...817..114Z}.

The instrumental magnitudes were then converted to those of the standard $BV$ (Vega) and $gri$ (AB) systems. The $BV$ calibration is based on the $BV$-band brightness of local stars from AAVSO Photometric All Sky Survey (APASS; \citealp{2009AAS...21440702H, 2010AAS...21547011H}), while $gri$ magnitudes are calibrated using reference stars from the Pan-STARRS catalogue \citep{2016arXiv161205242M}. In order to remove the host-galaxy light from the $gri$ data, we applied template subtraction to $gri$ images with Pan-STARRS archival images as template. Template subtraction was not performed for the $BV$ photometry, since the SN was sufficiently bright when those images were taken. The final flux-calibrated photometry is presented in Tables~\ref{Table: photo_LJT} and \ref{Table: photo_TNT}. 

The Ultra-Violet/Optical Telescope (UVOT; \citealp{2005SSRv..120...95R}) onboard the Neil Gehrels {\it Swift} Observatory started observations of SN 2018hti on 2018 November 8.52 and continued until around 140 d after discovery. The observations were made in the $uvw1$, $uvm2$, $uvw2$, $u$, $b$, and $v$ filters, and the data reduction follows that of the Swift Optical Ultraviolet Supernova Archive \cite[SOUSA;][]{2014Ap&SS.354...89B}. A 3\arcsec\ aperture is used to measure the source counts with an aperture correction based on an average point-spread function (PSF). The resultant UVOT photometry is listed in Table~\ref{Table: photo_Swift}. The {\it Swift} $b$- and $v$-band photometry are overall consistent (within the quoted uncertainties) with that of the LJT/TNT observations obtained in the Johnson $B$ and $V$ bands. We also include a set of $BVRI$ data observed on MJD 58715.98 by the AZT-22 1.5-m telescope (AZT) at Maidanak Astronomical Observatory \citep{2018NatAs...2..349E}. Flux-calibrated data, based on Pan-STARRS reference catalogue \citep{2016arXiv161205242M}, are given in Table~\ref{Table: photo_AZT}. We caution that the image subtraction has not been applied to the AZT data. So these points should be treated carefully or be considered as the upper limits of the brightness of SN 2018hti at that phase.

\begin{figure}
\center
\includegraphics[angle=0,width=1\textwidth]{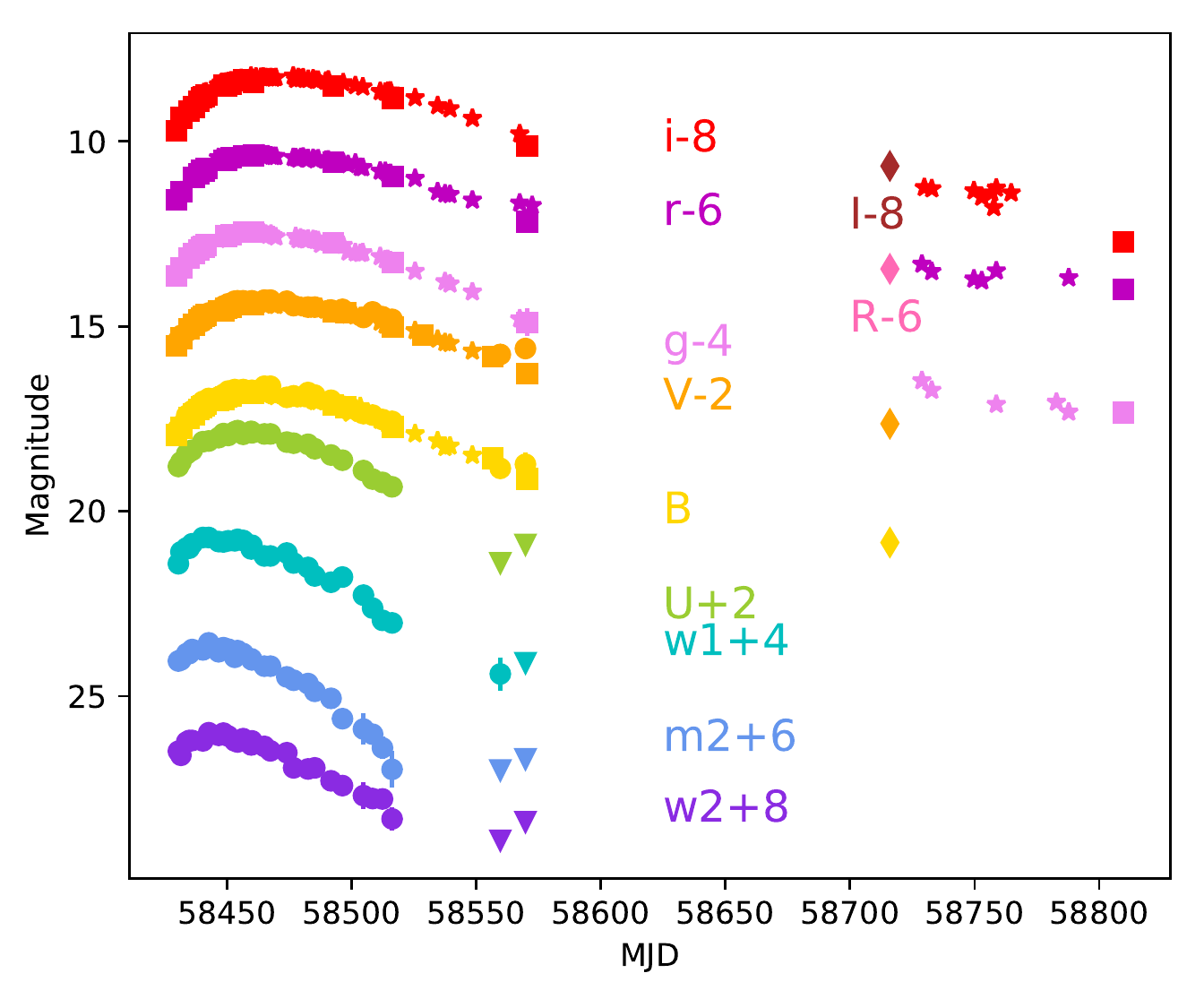}
\caption{Optical and ultraviolet light curves of SN 2018hti. Data are taken from the observations of LJT ($BVgri$; squares), TNT ($BVgri$; stars), AZT ($BVRI$; diamonds) and {\it Swift}/UVOT ($w2$, $m2$, $w1$, $u$, $b$, $v$; circles represent the observed Vega magnitudes and triangles show the lower limits).}
\label{fig: Photometry}
\end{figure}

All of the photometric data, spanning from $-34.8$ to +345.1 d relative to the epoch of $r$-band maximum, are displayed in Figure~\ref{fig: Photometry}. The maximum-light dates in $r$ and other bands are determined by fitting the observed data with low-order polynomials.

\subsection{Spectroscopy}
\label{SubSec: Spec}

A journal of the spectral observations of SN 2018hti is presented in Table \ref{Table: spec_Journal}. Our first spectrum was taken with LJT on 2018 November 7, five days after discovery. With the LJT+YFOSC system of the Yunan Astronomical Observatory (YNAO), we obtained a total of 10 spectra. We also collected spectra using the 2.16-m telescope at Xinglong Observatory (hereafter XLT) of NAOC, the Kast spectrograph \citep{1993Miller} on the 3-m Shane telescope at Lick Observatory (hereafter Lick), the Low-Resolution Imaging Spectrometer (LRIS; \citealp{1995PASP..107..375O}) on the Keck~I 10-m telescope (hereafter Keck), and the 3.5-m telescope at the Apache Point Observatory owned and operated by the Astrophysical Research Corporation (hereafter APO). The latest spectrum was taken by Keck $\sim 350$ d after discovery, which can be also used to diagnose the properties of the host galaxy because the SN was relatively faint at this phase. The spectra from the LJT were taken with the slit oriented along the parallactic angle \citep{fil82} to minimise the effects of atmospheric dispersion, while those from the XLT and APO were usually taken at small values of the airmass. The Keck/LRIS spectra were obtained with an atmospheric dispersion corrector.

IRAF routines were used to reduce the spectra obtained with the LJT, XLT, and APO. The fluxes of the spectra were calibrated based on the standard stars observed at similar airmasses on the same night, and telluric lines were also removed from most of the spectra. Using the extinction curves of local observatories, the spectra were further corrected for continuum atmospheric extinction. Standard procedures \citep[e.g.,][]{Silverman12} were used to extract and calibrate one-dimensional Keck/LRIS spectra from the two-dimensional CCD data; these included flat-fielding, cosmic-ray removal, optimal extraction \citep{horne86} and sky subtraction, flux calibration, and removal of telluric absorption through comparison with spectra of standard stars taken at similar airmass values.

\subsection{Spectral evolution}
\label{SubSec: SpecEvo}

\begin{figure}
\center
\includegraphics[angle=0,width=0.9\textwidth]{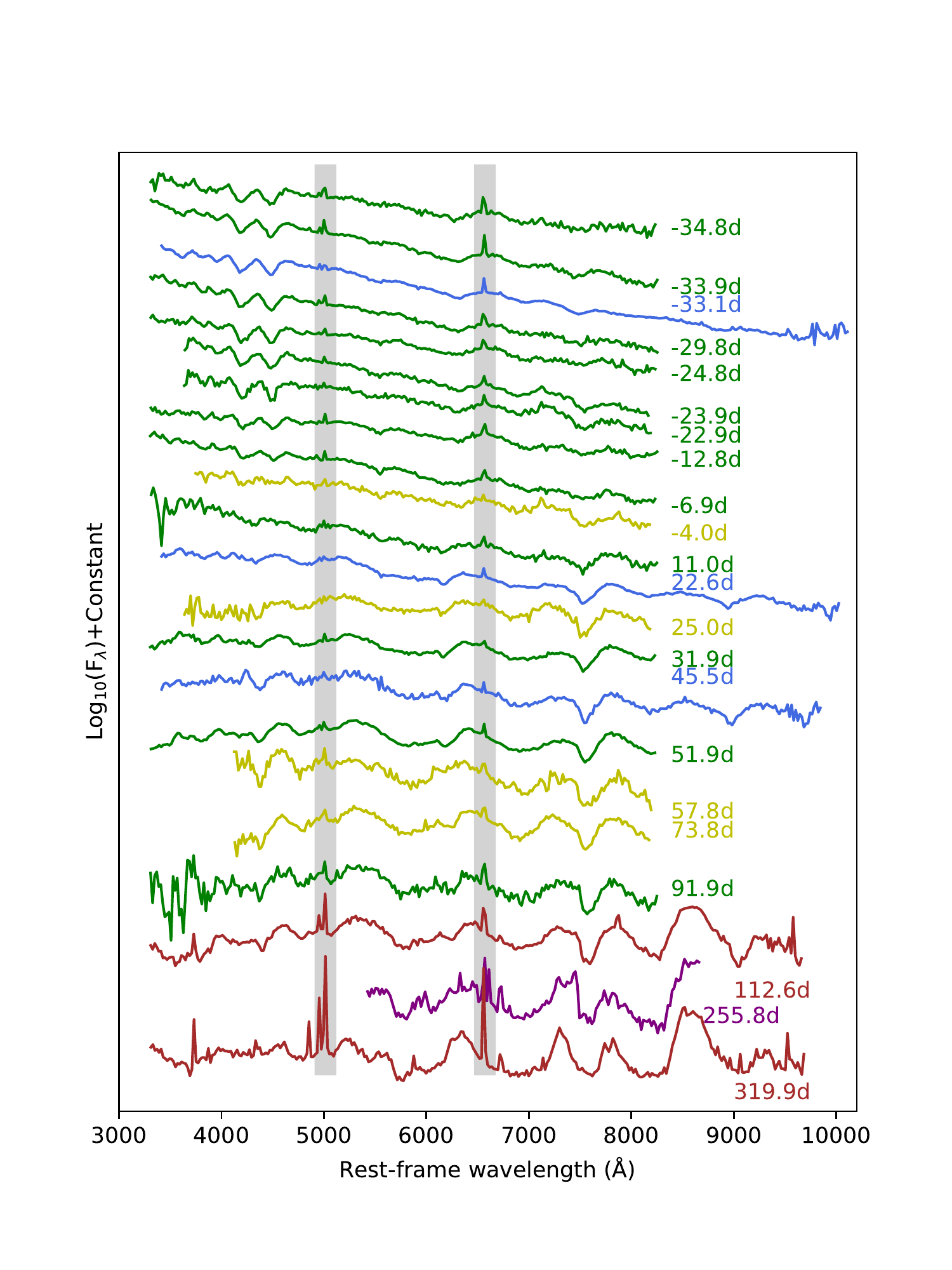}
\caption{Spectral evolution of SN 2018hti. Data were obtained by LJT (green), XLT  (yellow), APO (purple), Lick Shane telescope (blue), and Keck I telescope (brown). All spectra have been rebinned to 20\,\AA\ bin$^{-1}$.
  The grey shaded region highlights the H$\alpha$ and [O\,\textsc{iii}] lines from the host galaxy. The phases are marked in days relative to the epoch of $r$-band peak.}
\label{fig: Spec_all}
\end{figure}

A series of spectra of SN 2018hti is chronologically displayed in Figure~\ref{fig: Spec_all}. One can see that the early-time, pre-maximum spectra did not show significant evolution; they are characterized by blue continua, prominent multiple O\,\textsc{ii} absorption lines over the wavelength range 3500--5000\,\AA, and C\,\textsc{ii} and O\,\textsc{i} lines at longer wavelengths. These spectral features are similar to those of typical SLSNe~I at similar phases (e.g., \citealp{2016MNRAS.458.3455M, 2016ApJ...826...39N, 2018ApJ...855....2Q, 2019ApJ...882..102G}). At around maximum light, the major features at short wavelengths, such as the O\,\textsc{ii} blends, become gradually weak or even disappear, likely due to the expansion of the ejecta and hence the cooling of the photosphere. After maximum light, the continuum further becomes redder as a result of a continuous decrease in photospheric temperature. In the nebular phase, the forbidden lines of [Fe\,\textsc{ii}], [O\,\textsc{i}], and [Ca\,\textsc{ii}], gradually dominate the spectra.

\begin{figure}
\center
\includegraphics[angle=0,width=1\textwidth]{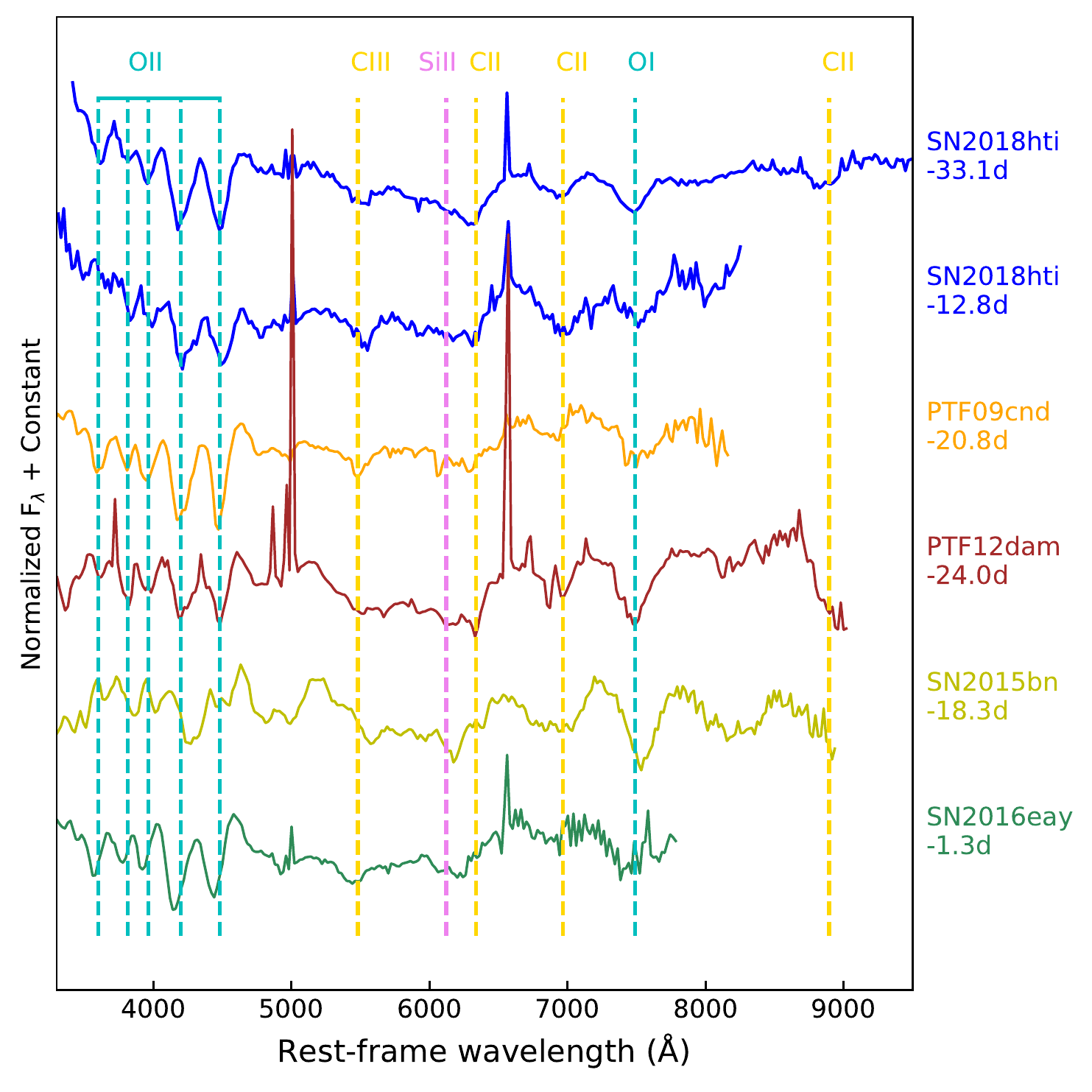}
\caption{Comparison of early-time spectra of SN 2018hti with well-observed SLSNe~I (PTF09cnd, PTF12dam, SN 2015bn, and SN 2016eay). All spectra have been rebinned to 20\,\AA\ bin$^{-1}$. The dashed vertical lines mark the identifications of O\,\textsc{i}, O\,\textsc{ii}, Si\,\textsc{ii}, C\,\textsc{ii}, and C\,\textsc{iii} that have been blueshifted by a velocity of 11000\,km\,s$^{-1}$. Data references: PTF09cnd \citep{2011Natur.474..487Q}, PTF12dam \citep{2013Natur.502..346N}, SN 2015bn \citep{2016ApJ...826...39N}, and SN 2016eay \citep{2017ApJ...835L...8N}; these data are retrieved from the open supernova catalogue \citep{2017ApJ...835...64G}.}
\label{fig: Spec_1}
\end{figure}

\begin{figure}
\center
\includegraphics[angle=0,width=1\textwidth]{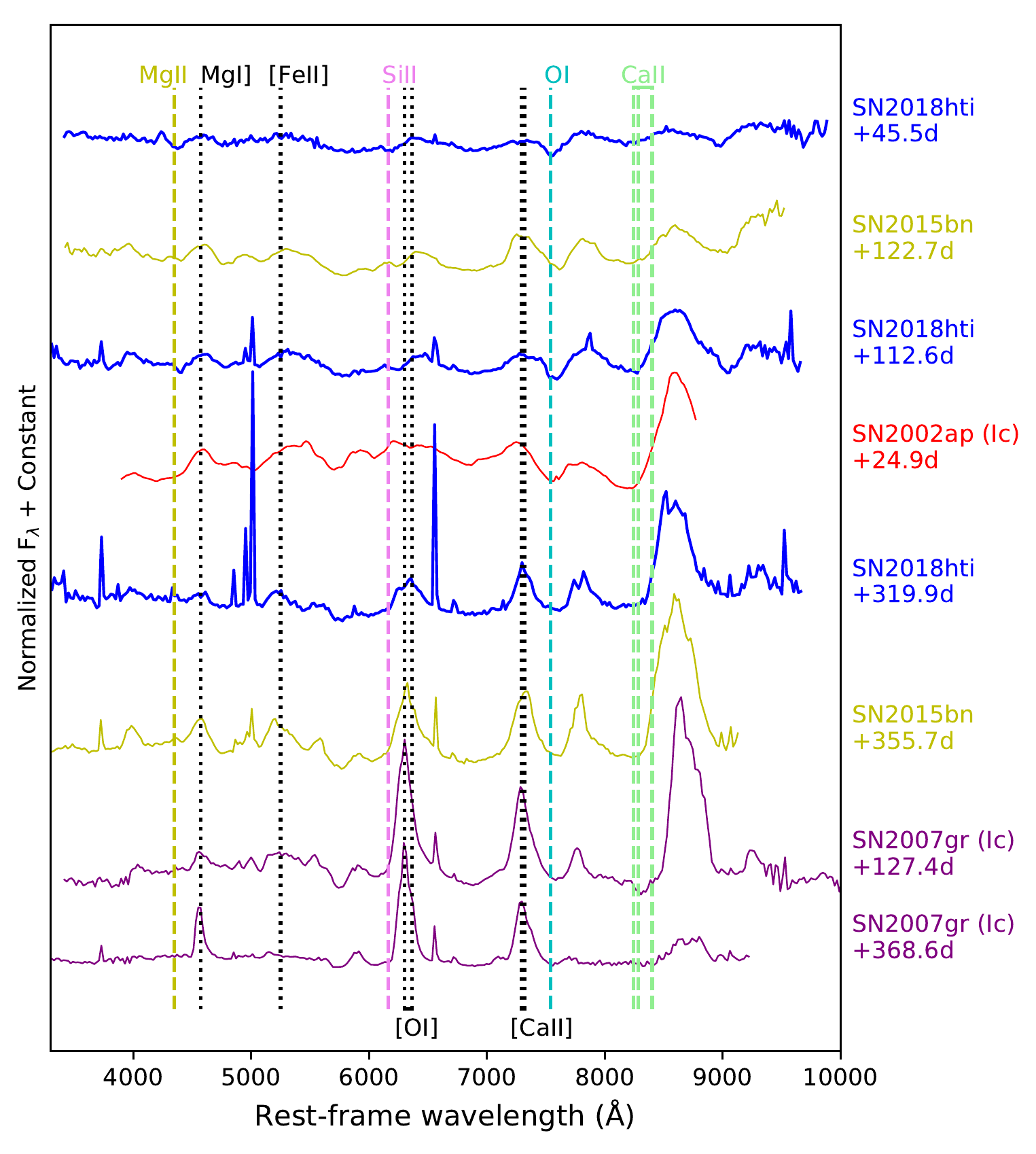}
\caption{Comparison of the post-peak spectra of SN 2018hti and a well-observed SLSN~I (SN 2015bn).  All spectra have been rebinned to 20\,\AA\ bin$^{-1}$. The dashed vertical lines mark the location of O\,\textsc{i}, Si\,\textsc{ii}, Mg\,\textsc{ii}, and Ca\,\textsc{ii} lines that have been blueshifted by a velocity of 8000\,km\,s$^{-1}$. The dotted vertical lines show the emission lines ([O\,\textsc{i}], Mg\,\textsc{i}], [Ca\,\textsc{ii}], and [Fe\,\textsc{ii}]) without a blueshift. For comparison, spectra of two well-observed SNe~Ic (SN 2002ap and SN 2007gr) are displayed. Data references: SN 2002ap \citep{2003PASP..115.1220F}, SN 2007gr \citep{2019MNRAS.482.1545S}, and SN 2015bn \citep{2016ApJ...826...39N, 2017ApJ...835...13J}; these data are retrieved from the open supernova catalogue \citep{2017ApJ...835...64G}.}
\label{fig: Spec_3}
\end{figure}

Figures~\ref{fig: Spec_1}-\ref{fig: Spec_3} show a comparison of the spectra with those of some other well-observed SLSNe~I at two different evolutionary stages. All spectra are normalized to their corresponding continua in order to highlight the spectral features. From comparison of pre-maximum spectra, we find that SN 2018hti is similar to other SLSNe~I like PTF09cnd and PTF12dam, all showing the strong O\,\textsc{ii} absorption complex, which may be blended with Ca\,\textsc{ii}, Fe\,\textsc{iii}, and ionized carbon lines (e.g., \citealp{2016ApJ...826...39N, 2019ApJ...882..102G}). To explain this absorption complex, \citet{2018ApJ...855....2Q} propose an absorption model and construct five O\,\textsc{ii} features with effective wavelengths of 4650.71\,\AA, 4357.97\,\AA, 4115.17\,\AA, 3959.83\,\AA,  and 3737.59\,\AA. Besides the pronounced O\,\textsc{ii} lines, other main features in the early-time spectra of SLSNe~I include C\,\textsc{iii} $\lambda5690$, C\,\textsc{ii} $\lambda7234$, and O\,\textsc{i} $\lambda7774$ (e.g., \citealp{2017ApJ...840...57Y}). Note that the wide absorption minimum near 6500\,\AA\ in the spectra of SN 2018hti can be attributed to C\,\textsc{ii} $\lambda6580$. This carbon feature and Si\,\textsc{ii} $\lambda6355$ can be clearly detected in the spectra of PTF2012dam and SN 2015bn (see also \citealp{2016ApJ...826...39N, 2018ApJ...855....2Q}). Inspecting the early-time spectra indicates that Si\,\textsc{ii} $\lambda6355$ might also exist, though not as strong as C\,\textsc{ii} $\lambda6580$ in SN 2018hti and SN 2016eay.

By comparing the effective wavelengths suggested by \citet{2018ApJ...855....2Q} with the observed minima of the O\,\textsc{ii} absorption lines in the spectrum taken at $t \approx -33$ d, we estimate a blueshifted velocity of about 11000 km s$^{-1}$ for SN 2018hti, consistent with that inferred from the spectra of PTF09cnd and PTF12dam obtained at similar phases. In comparison, SN 2015bn seems to have a slower velocity of O\,\textsc{ii} ($\sim8000$ km s$^{-1}$) while SN 2016eay has a larger value ($\sim13000$ km s$^{-1}$). 

Figure~\ref{fig: Spec_3} shows the post-peak spectra\footnote{Here, all absorption lines are marked assuming a blueshifted velocity of 8000 km s$^{-1}$. Following \citep{2016ApJ...826...39N}, we set the [O\,\textsc{i}], Mg\,\textsc{i}], and [Ca\,\textsc{ii}] with zero velocity. It is reasonable to assume that those emission lines form deep inside the ejecta and thus have a lower blueshifted velocity. Based on similar thinking, [Fe\,\textsc{ii}] is shown without a blueshift. When it comes to the nebular phase, all nebular lines should be at low velocity.}. After maximum light, SN 2018hti exhibits slow spectral evolution, similar to that of the comparison SLSNe~I. The absorption features between 4000\,\AA\ and 5000\,\AA\ can be attributed to Fe\,\textsc{ii} and Mg\,\textsc{ii}, as identified in some SLSNe~I (e.g., SN 2016eay; \citealp{2016ApJ...826...39N}) and SNe~Ic (e.g., SN 2017ein; \citealp{2019ApJ...871..176X}). The absorption features coincident with the C\,\textsc{iii} $\lambda5690$ and C\,\textsc{ii} $\lambda7234$ lines become wide absorption troughs. The C\,\textsc{ii} $\lambda6580$ line gradually weakens and disappears, while the Si\,\textsc{ii} $\lambda6355$ absorption starts to dominate the wavelength near $\sim6200$\,\AA\ until the [O\,\textsc{i}] $\lambda\lambda$6300, 6364 blend becomes strong. Redward of this feature, one can see the prominent O\,\textsc{i} $\lambda7774$ absorption, which may blend with Mg\,\textsc{ii} $\lambda\lambda7877$, 7896 \citep{2019ApJ...871..102N}. The absorption feature near $\sim9000$\,\AA\ in the early-time spectra was likely related to C\,\textsc{ii} $\lambda9234$. Considering that the other C lines are relatively weak, this feature can also be a blend of O\,\textsc{i} and Mg\,\textsc{ii}, which is used to explain the spectra of SN 2017egm \citep{2018ApJ...853...57B}.

After a few weeks past maximum brightness, the Ca\,\textsc{ii} near-infrared (NIR) triplet emerges in the spectra and gradually gains strength. In the nebular phase, the photospheric recession and cooling of the ejecta lead to the recombination of ionized Ca, shown as the Ca\,\textsc{ii} NIR triplet and [Ca\,\textsc{ii}] near 7300\,\AA\ in the $t \approx 319$ d Keck spectrum. This spectrum also displays emission lines coincident with Mg\,\textsc{i}] $\lambda4571$ and [Fe\,\textsc{ii}] $\lambda5250$. These features are similarly seen in the nebular spectra of some SLSNe~I (e.g., SN 2015bn; \citealp{2017ApJ...835...13J}) and SNe~Ic (e.g., SN 2002ap and SN 2007gr; \citealp{2003PASP..115.1220F, 2010MNRAS.408..947M, 2014ApJ...790..120C}). Note that the [Fe\,\textsc{ii}] $\lambda5250$ feature can be produced by a blend of [Fe\,\textsc{ii}] $\lambda5250$, Mg\,\textsc{i} $\lambda5180$, and [O\,\textsc{i}] $\lambda5577$ \citep{2017MNRAS.468.4642I, 2017ApJ...835...13J}. In contrast to SN 2010md \citep{2018ApJ...855....2Q}, there is no clear evidence for the presence of He in SN 2018hti.

In common with other SLSNe~I (e.g., \citealp{2010ApJ...724L..16P, 2016ApJ...828L..18N, 2017ApJ...845...85L, 2019ApJ...872...90B}), SN 2018hti displays post-maximum spectra quite similar to those of SNe~Ic except that the latter evolve faster. The $t \approx 112.6$ d spectrum of SN 2018hti resembles that of SN 2002ap taken at $t \approx 24.9$ d, both characterized by prominent absorption features of O, Mg, Fe, and Ca. After $t \approx 100$ d, SNe~Ic usually exhibit nebular lines, such as [O\,\textsc{i}], [Ca\,\textsc{ii}], and Ca\,\textsc{ii}, as seen in the $t \approx 127$ d spectrum of SN 2007gr (see also \citealp{2003PASP..115.1220F, 2010MNRAS.408..947M}). Similar features are not observed in the $\sim 100$ d spectrum of SN 2018hti but appear in its $t \gtrsim 300$ d spectrum and that of SN 2015bn at a similar phase. In comparison, the narrow [O\,\textsc{i}] and Mg\,\textsc{i}] lines seen in SN 2007gr are not visible in SN 2018hti at $t = 319$ d, while the O\,\textsc{i} and Ca\,\textsc{ii} lines have weakened and even disappeared in SN 2007gr at such a late phase. Overall, the spectral evolution of SN 2018hti and other SLSNe~I seems to be a factor of two slower than that of SNe~Ic. The slow spectral evolution could be attributed to larger ejecta mass and the existence of persistent injection energy.

Overall, the spectra of SN 2018hti are similar to those of other SLSNe~I; the pronounced features at 6000--8000\,\AA\ allow us to categorize SN 2018hti as a PTF12dam-like event \citep{2018ApJ...855....2Q}. \citet{2015ApJ...814..108Y, 2017ApJ...848....6Y} reported the presence of late-time H$\alpha$ emission for three SLSNe~I, suggesting interaction between SN ejecta and the H-rich shell at a distance of $\sim 10^{16}$ cm. Assuming an ejecta velocity of 11000 km s$^{-1}$ and a similar distance for the hydrogen-rich CSM shell, H$\alpha$ emission is expected to emerge in the $t \sim 100$ d spectrum, but this was not observed in SN 2018hti.

\subsection{The host galaxy}
\label{SubSec: host}
SLSNe~I are generally found to be harboured by low-luminosity dwarf galaxies (e.g., \citealp{2014ApJ...787..138L, 2015MNRAS.449..917L, 2016ApJ...830...13P}). This trend seems to hold true with SN 2018hti, for which the host galaxy is very faint and nearly invisible in the TNT and LJT images. This is partially due to the large line-of-sight reddening from the Milky Way. In the deeper Pan-STARRS images, an isolated and compact host galaxy can be distinguished near the location of SN 2018hti (see Figure~\ref{fig: FinderChart}). The SN is 0.39$\arcsec$ north and 0.34$\arcsec$ east of the host nucleus. The host galaxy is measured to have apparent magnitudes of $20.93\pm0.05$ in $g$ and $20.33\pm0.04$ in $r$,  corresponding to absolute magnitudes of $M_g=-17.75$ and $M_r=-17.91$ after correction for Galactic extinction. It is dimmer than the hosts of most core-collapse SNe and long-duration gamma-ray bursts, both of which are preferentially found to be brighter than $-18.0$ mag in $B/g$ band (e.g., \citealp{2010ApJ...721..777A, 2014ApJ...787..138L}).

\begin{figure}
\center
\includegraphics[angle=0,width=1\textwidth]{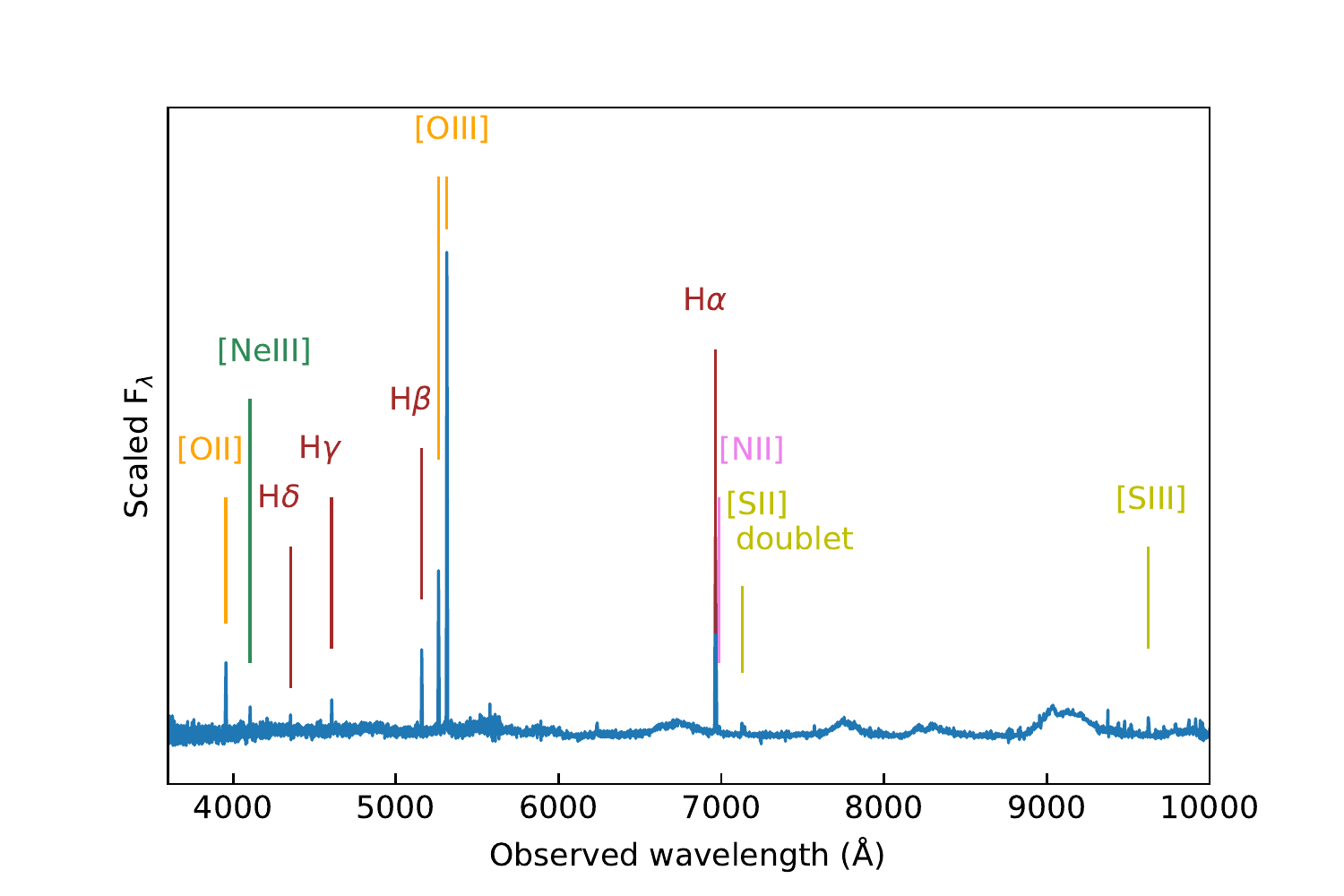}
\caption{Host-galaxy emission lines of SN 2018hti identified in the Keck spectrum obtained on 2019 October 28. Each bin is 0.5--0.6\,\AA.}
\label{fig: host_lines}
\end{figure}

Despite the low surface brightness, emission lines from the host galaxy of SN 2018hti can clearly be seen in the SN spectra, especially the $t \approx 319$ d spectrum taken by the Keck telescope, as shown in Figure~\ref{fig: host_lines}. The majority of host emission lines originate from  neutral hydrogen and low-ionization elements such as N, O, Ne, and S. We fit the line profiles with a Gaussian function and the continuum with a linear function. As the [N\,\textsc{ii}] $\lambda$6584 line is close to H$\alpha$, we apply a simultaneous fit to these two features. (The [N\,\textsc{ii}] $\lambda$6548  line was too weak to measure.) A simultaneous fit is also applied to the [S\,\textsc{ii}] doublet. From these narrow emission lines, we measure a redshift of $z=0.0612\pm0.0001$ for the host galaxy. We also estimate the observed flux, equivalent width (EW), and full width at half-maximum intensity (FWHM) of each individual emission line from the host galaxy; the results are presented in Table~\ref{Table: host_lines_fit}. We caution that measurements of the above parameters might deviate from the intrinsic values as they are obtained from the spectrum characterized by the nebular features of the SN. However, the observed flux ratios, which are used below, should be accurate because the host-galaxy emission lines are very strong.

Previous studies (e.g., \citealp{2004MNRAS.348L..59P, 2008A&A...488..463M, 2017MNRAS.465.1384C}) proposed that the intensity ratios of strong emission lines can be good indicators of the oxygen abundance of a galaxy. These flux ratios provide empirical calibrations of oxygen abundance in the form of polynomial functions. In our analysis, we calculate the flux ratios between some strong lines, including R2 ([O\,\textsc{ii}] $\lambda$3727/H$\beta$), R3 ([O\,\textsc{iii}] $\lambda$5007/H$\beta$), R23 (([O\,\textsc{ii}] $\lambda$3727 + [O\,\textsc{iii}] $\lambda$4959, 5007)/H$\beta$), O32 ([O\,\textsc{iii}] $\lambda$5007/[O\,\textsc{ii}] $\lambda$3727), N2 ([N\,\textsc{ii}] $\lambda$6584/H$\alpha$), and O3N2 (([O\,\textsc{iii}] $\lambda$5007/H$\beta$)/([N\,\textsc{ii}] $\lambda$6584/H$\alpha$)). Based on the empirical relations obtained by \citet{2017MNRAS.465.1384C}, the abundance parameter 12 + log(O/H) is estimated to be 8.21, 8.18, 8.13, and 8.13 from the indicators R2, O32, N2, and O3N2, respectively. The values derived from the R3 and R23 ratios, which are both out of the range of applicability (see Table 2 in \citealp{2017MNRAS.465.1384C}), are not presented here. Thus, an average value of 12 + log(O/H) = 8.16 is adopted as the oxygen abundance for the host galaxy of SN 2018hti, which corresponds to 0.3\,$Z_\odot$ (assuming a solar oxygen abundance of 8.69; \citealp{2009ARA&A..47..481A}). This value is consistent with previous results that most SLSNe~I reside in low-metallicity galaxies \citep{2014ApJ...787..138L, 2016ApJ...830...13P, 2018MNRAS.473.1258S}. Nevertheless, we caution that the line diagnostics mentioned above might be affected by dust, ionization properties, and relative elemental abundances (see \citealp{2019A&ARv..27....3M} for an extensive discussion).

The intrinsic Balmer-line ratio is given by $H\alpha/H\beta_\mathrm{int}=2.86$, in the context of the Case B recombination model \citep{1989agna.book.....O, 2013AJ....145...47M}. This Balmer-line ratio can thus be used to estimate the host-galaxy reddening. For SN 2018hti, the host-galaxy reddening is inferred to be $E(B - V) = 0.05$ mag. With this reddening, the derived oxygen abundance is modified to be 8.17. The star formation rates (SFRs) are found to be $0.31\pm0.01\,$ M$_\odot$ yr$^{-1}$ from the H$\alpha$ luminosity and $0.34\pm0.02\,$M$_\odot$ yr$^{-1}$ from the [O\,\textsc{ii}] luminosity \citep{1998ARA&A..36..189K}. Considering the uncertainties in the above empirical relations, the two derived SFRs are consistent with each other.

\begin{figure}
\center
\includegraphics[angle=0,width=1\textwidth]{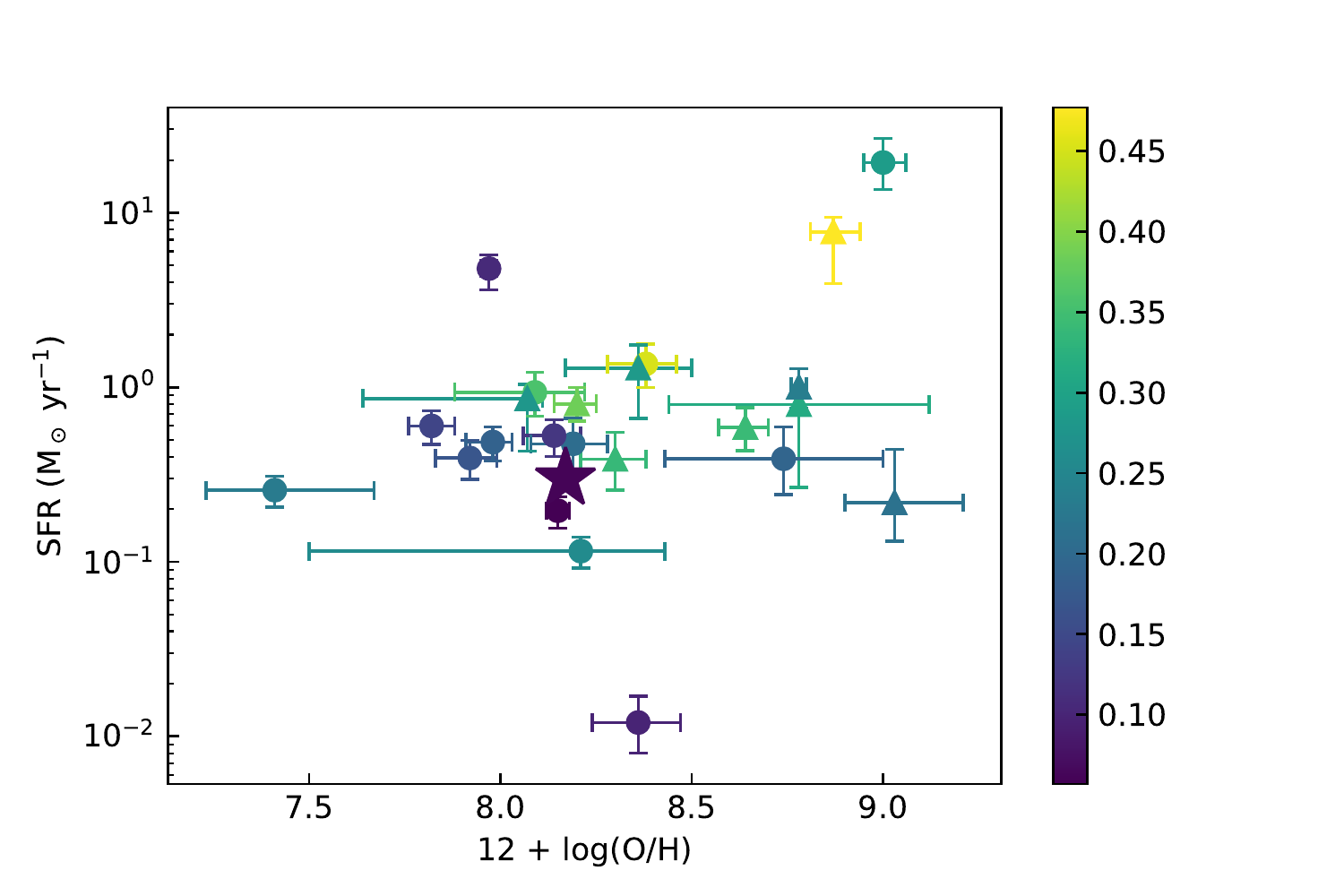}
\caption{The metallicity versus star formation rate derived for SN 2018hti (star) and other SLSNe~I (circle) and SLSNe~II (triangle) available from \citet{2016ApJ...830...13P}. The colours indicate the redshifts.}
\label{fig: host_SFRMetal}
\end{figure}

We conclude that SN 2018hti occurred in a faint galaxy with a low metallicity and a SFR of $0.3\,$M$_\odot$ yr$^{-1}$. Although SLSNe are found in galaxies covering a wide range of SFR and metallicity, most of them tend to reside in galaxies with SFR $\approx 0.1$--2\,M$_\odot$ yr$^{-1}$ (see also Figure~\ref{fig: host_SFRMetal}). Among the current sample of SLSNe, the Type I subclass occurs preferentially in dwarf galaxies with $\lesssim 0.5$ solar metallicity while the Type II subclass does not show such a tendency (e.g., \citealp{2014ApJ...787..138L, 2015MNRAS.449..917L, 2016ApJ...830...13P, 2017MNRAS.470.3566C}). Therefore, the host galaxy of SN 2018hti shares properties similar to those of other SLSNe~I.

\section{Light Curve}
\label{Sec: Photometry}

\subsection{Photometric evolution}
\label{SubSec: PhotEvo}

\begin{figure}
\center
\includegraphics[angle=0,width=1\textwidth]{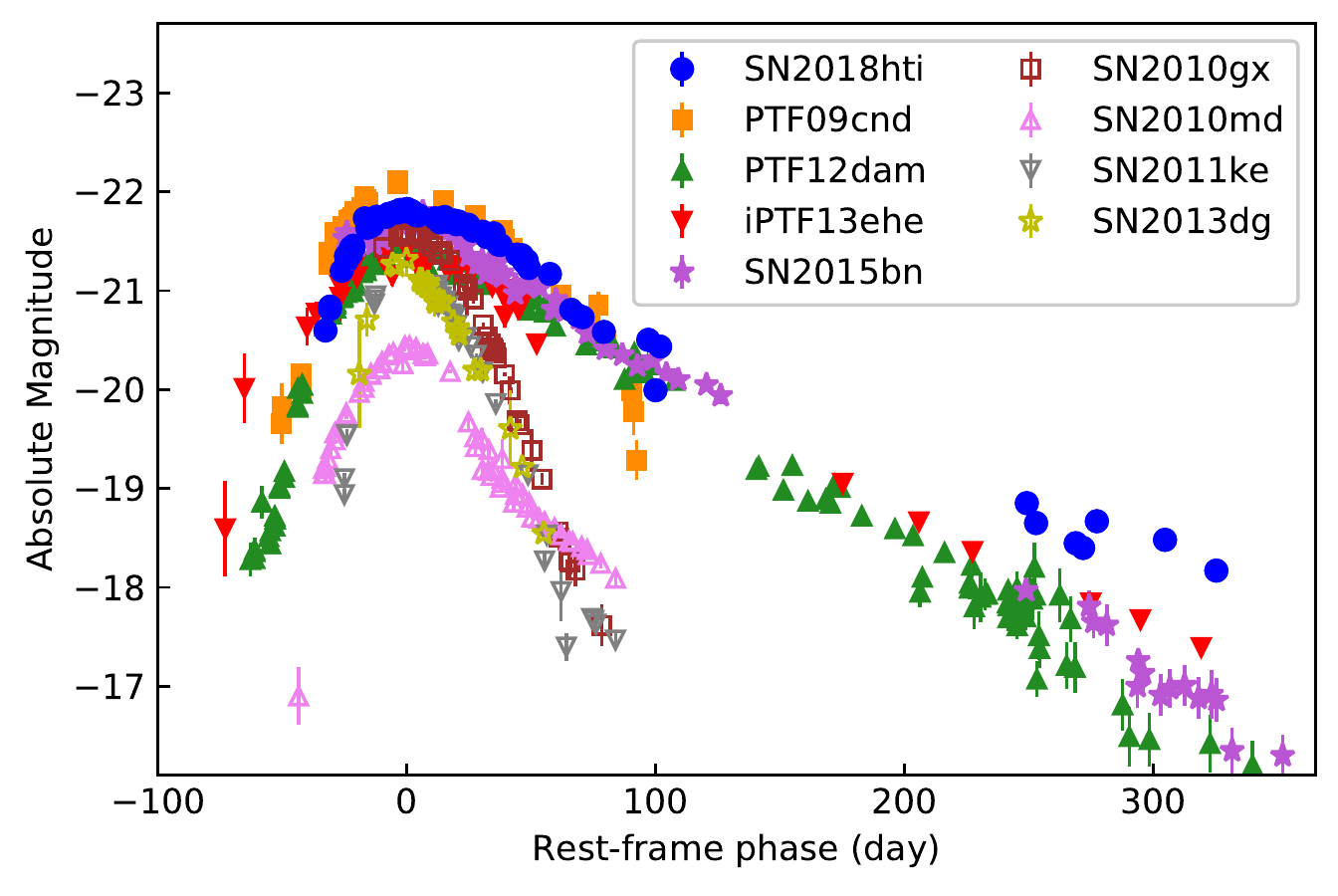}
\caption{The $r$-band light curves of SN 2018hti (blue circles), slowly declining SLSNe~I (PTF09cnd, PTF12dam, iPTF13ehe, SN 2015bn; filled markers), and rapidly declining SLSNe~I (SN 2010gx, SN 2010md, SN 2011ke, SN 2013dg; empty markers). The curves are corrected for Galactic extinction and shown in rest-frame days relative to the epoch of the optical peak. Data references: PTF09cnd \citep{2018ApJ...860..100D}, SN 2010gx \citep{2010ApJ...724L..16P}, SN 2010md \citep{2013ApJ...770..128I, 2018ApJ...860..100D}, SN 2011ke \citep{2013ApJ...770..128I, 2018ApJ...860..100D}, PTF12dam \citep{2013Natur.502..346N, 2017ApJ...835...64G}, iPTF13ehe \citep{2015ApJ...814..108Y}, SN 2013dg \citep{2014MNRAS.444.2096N}, and SN 2015bn \citep{2016ApJ...828L..18N, 2016ApJ...826...39N}; these data are retrieved from the open supernova catalogue \citep{2017ApJ...835...64G}.}
\label{fig: comp_r}
\end{figure}

Figure~\ref{fig: comp_r} displays the evolution of the optical light curves of SN 2018hti and a comparison with some representative SLSNe~I, including PTF09cnd, SN 2010gx, SN 2010md, SN 2011ke, PTF12dam, iPTF13ehe, SN 2013dg, and SN 2015bn. These comparison SLSNe~I can be divided into slowly and rapidly evolving subclasses that are clearly separated from each other in terms of post-maximum decline rates, though it has been argued that a larger sample of SLSNe~I may not follow such a bimodal distribution \citep{2017ApJ...850...55N, 2018ApJ...860..100D, 2018ApJ...852...81L}.

We perform a linear fit to their light curves within rest-frame 50 d after peak brightness. The slowly evolving SLSNe~I such as PTF09cnd, PTF12dam, iPTF13ehe, and SN 2015bn have a post-peak decay rate of $<0.02$ mag d$^{-2}$, while the rapidly evolving sample (i.e., SN 2010gx, SN 2010md, SN 2011ke, and SN 2013dg) have a decline rate of 0.03--0.05 mag d$^{-2}$. SLSNe~I with slower post-peak decay rates tend to have longer rise times \citep{2015MNRAS.452.3869N}, except SN 2010md which rises slowly but decays at a faster rate. For SN 2018hti, the post-maximum decline rate (within 50 d) is measured to be 0.01 mag d$^{-1}$, putting it into the slowly declining subclass. After $t = 200$ d from the peak, SN 2018hti shows a slow luminosity evolution, similar to iPTF13ehe, PTF12dam, and SN 2015bn, suggesting that they may have a similar power source at this phase. Further subdivision within SLSNe~I is also supported by the spectroscopic analysis. \citet{2018ApJ...855....2Q} established the presence of two subgroups of SLSNe~I based on the spectral similarity to either PTF12dam or SN 2011ke, with the latter having faster spectral evolution. Among the comparison sample, SN 2018hti shows spectral evolution similar to that of PTF12dam, consistent with the result from the photometric evolution.

\begin{figure}
\center
\includegraphics[angle=0,width=0.8\textwidth]{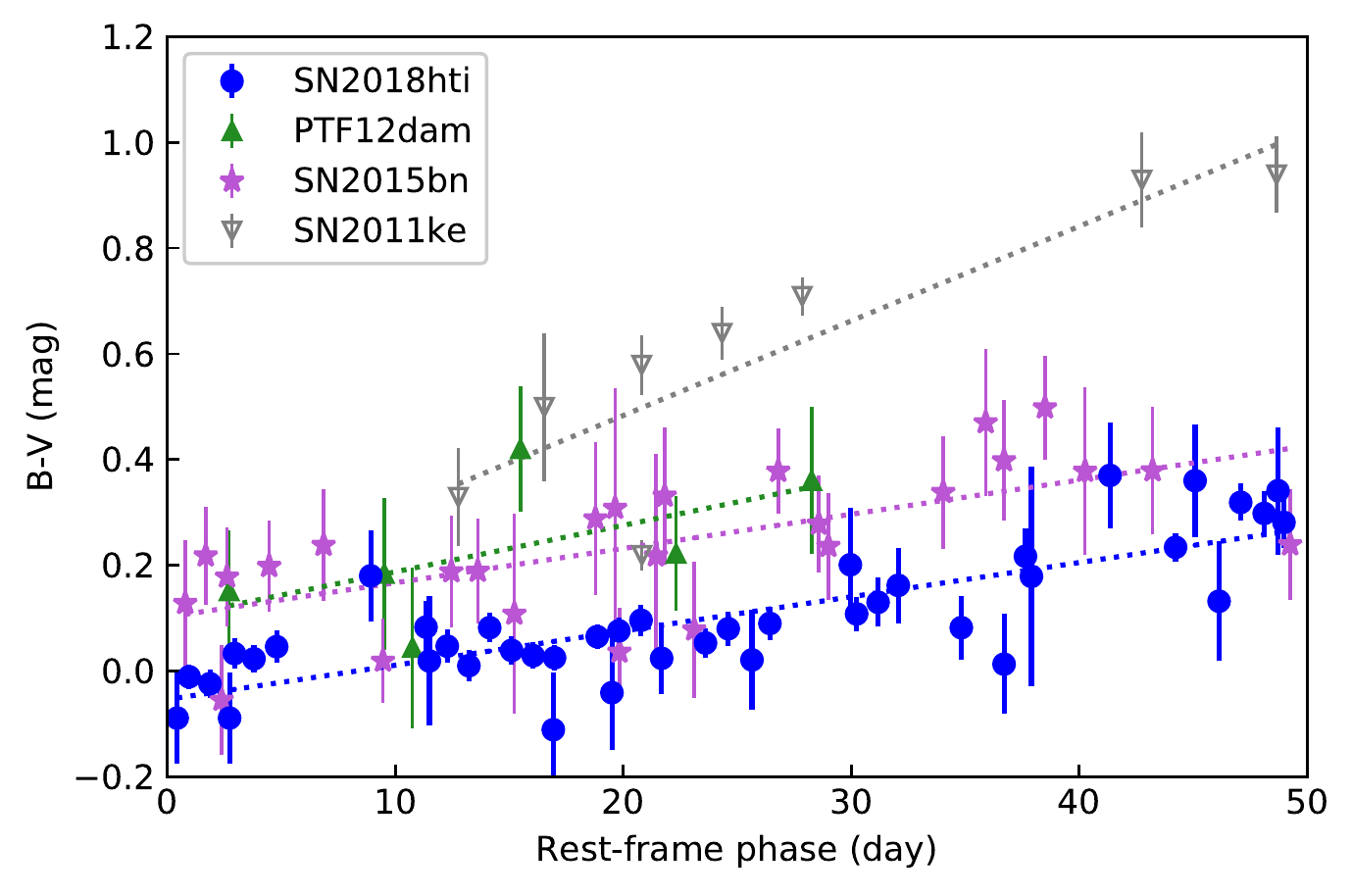}
\includegraphics[angle=0,width=0.8\textwidth]{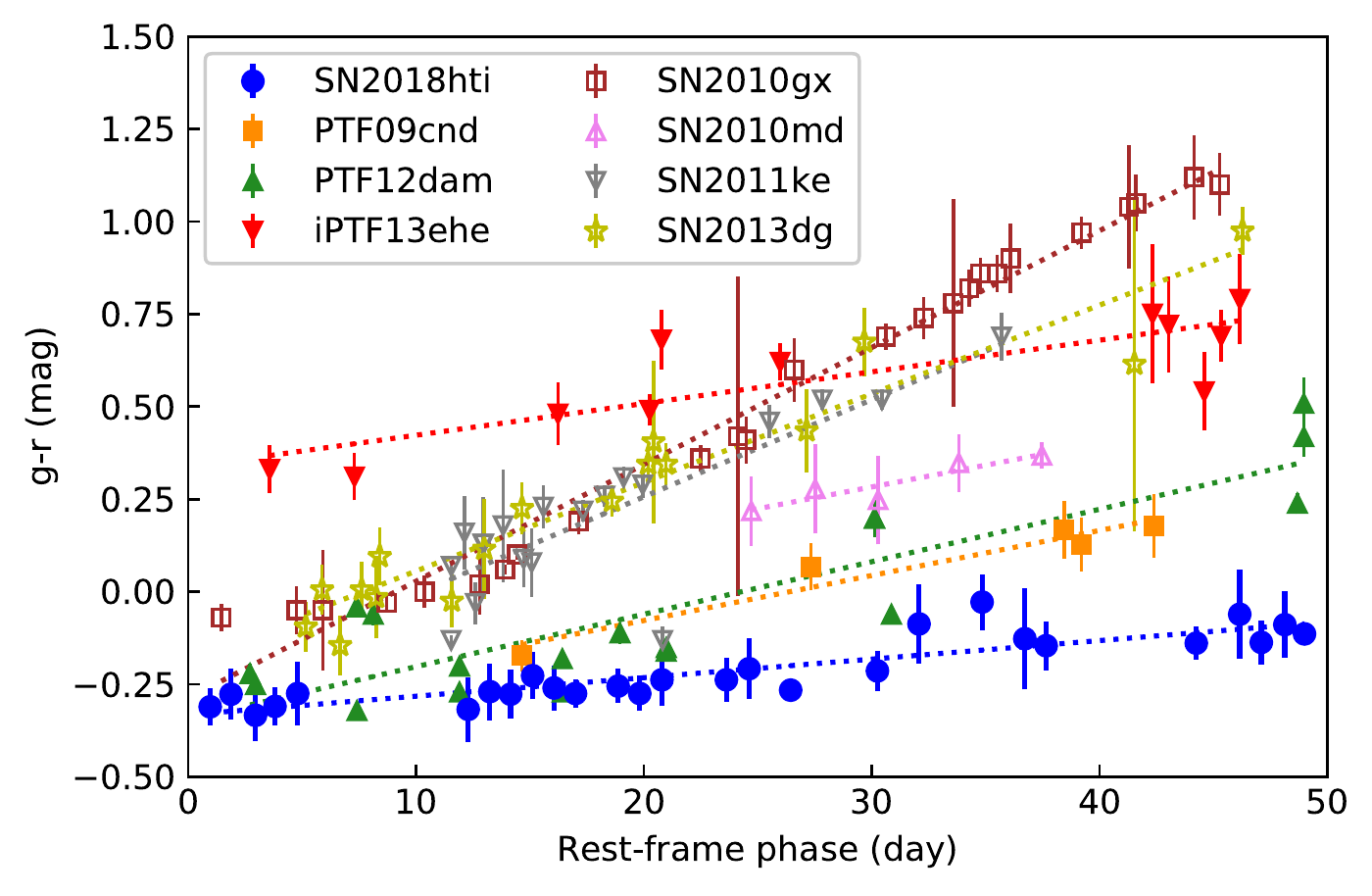}
\caption{The $B-V$ (top) and $g - r$ (bottom) evolution of SN 2018hti (blue circles) and two subclasses of SLSNe~I (filled markers for the slowly declining and empty markers for the rapidly declining ones). The colour curves are corrected for Galactic extinction and shown in rest-frame days relative to the epoch of the optical peak. The dotted lines represent the best linear fits to the colour evolution. Data references: PTF09cnd \citep{2018ApJ...860..100D}, SN 2010gx \citep{2010ApJ...724L..16P}, SN 2010md \citep{2013ApJ...770..128I, 2018ApJ...860..100D}, SN 2011ke \citep{2013ApJ...770..128I, 2018ApJ...860..100D}, PTF12dam \citep{2013Natur.502..346N, 2014Ap&SS.354...89B, 2017ApJ...835...64G}, iPTF13ehe \citep{2015ApJ...814..108Y}, SN 2013dg \citep{2014MNRAS.444.2096N}, and SN 2015bn \citep{2016ApJ...826...39N}; these data are retrieved from the open supernova catalogue \citep{2017ApJ...835...64G}.}
\label{fig: comp_color}
\end{figure}

Figure~\ref{fig: comp_color} shows the colour evolution of SN 2018hti. The $B - V$ colour of SN 2018hti evolves redward at a rate of $\sim 0.006$ mag d$^{-1}$), in rough agreement with the slow rate derived from SN 2015bn and PTF12dam. According to the overall evolution of the $B-V$ colour, SN 2018hti is bluer than the above two slowly evolving SLSNe~I for the first $\lesssim50$ d after peak brightness. In comparison, SN 2011ke is characterized by rapid redward evolution after maximum light --- at a rate of $0.018$ mag d$^{-1}$, though both slowly and rapidly declining subtypes cluster around $-0.1$--0.3 mag near maximum light. At $t \lesssim 50$ d from the peak, this rapidly evolving SLSN~I cooled down and reddened to $B - V \approx 0.9$ mag, in contrast to $\sim0.3$ mag for SN 2018hti.

The evolution of the $g - r$ colour is shown in the bottom panel of Figure~\ref{fig: comp_color}. Among the comparison sample, SN 2010gx, SN 2011ke, and SN 2013dg have large slopes of colour evolution (0.02--0.03 mag d$^{-1}$), quite distinct from the slowly evolving sample, such as iPTF13ehe (0.009 mag d$^{-1}$), PTF09cnd (0.012 mag d$^{-1}$), PTF12dam (0.014 mag d$^{-1}$), and SN 2018hti (0.005 mag d$^{-1}$). We notice that the $g - r$ colour of SN 2018hti shows the much slower redward evolution than all of the comparison SLSNe~I. This suggests the existence of continuous input of energy in SN 2018hti during its long-timescale expansion and cooling phase.

\begin{figure}
\center
\includegraphics[angle=0,width=0.8\textwidth]{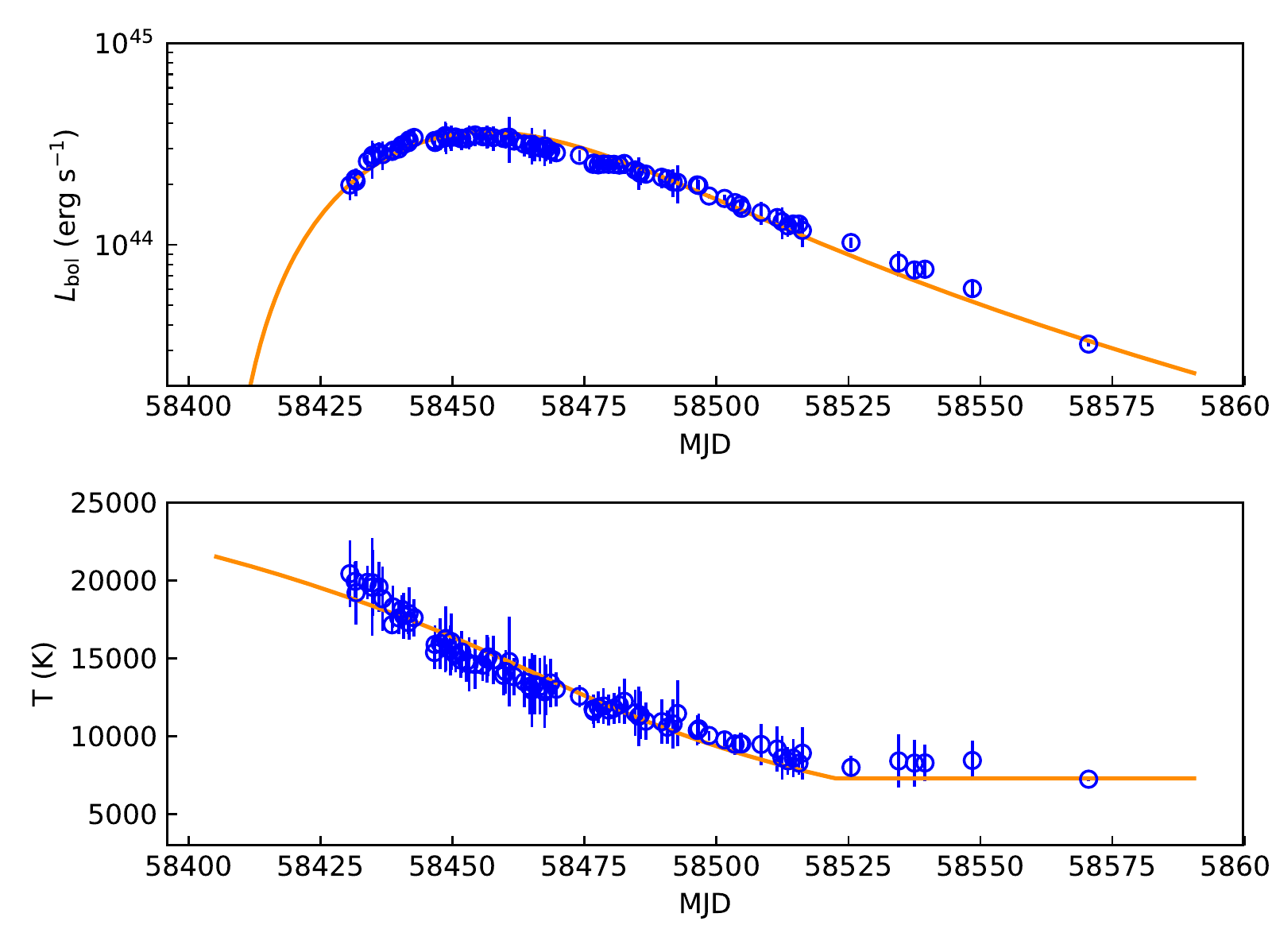}
\caption{Evolution of bolometric light curve (upper) and temperature (bottom) of SN 2018hti. The solid lines represent the theoretical curves from the magnetar-powered model.}
\label{fig: LT_fit}
\end{figure}

Applying the absorbed blackbody model \citep{2017ApJ...850...55N}\footnote{Significant absorption is commonly observed in the ultraviolet part of early-phase SLSNe~I SED (e.g., \citealp{2017ApJ...850...55N}), and spectra which exhibit strong absorption features of ionized carbon, silicon, magnesium, and titanium at $\sim 2000$--3000\,\AA\ (e.g., \citealp{2011ApJ...743..114C, 2011Natur.474..487Q, 2016MNRAS.458.3455M, 2017ApJ...835L...8N, 2018ApJ...855....2Q}). \citet{2017ApJ...850...55N} describe such an SED as a modified blackbody with flux suppression at $\lambda< 3000$\,\AA.} to fit the spectral energy distribution (SED) constructed from the extinction-corrected $UBVgri$-band photometry of SN 2018hti, we derived the bolometric luminosity ($L_\mathrm{bol}$) and effective temperature ($T_\mathrm{eff}$). The evolution of $L_\mathrm{bol}$ and $T_\mathrm{eff}$ are shown in Figure~\ref{fig: LT_fit}. The peak luminosity of SN 2018hti is calculated as $\sim3.5\times10^{44}$ erg s$^{-1}$, with the emission peak appearing earlier at shorter wavelengths. One can see that the ejecta are as hot as $\sim 13000$~K at $r$-band maximum light and have a temperature of $\sim15000$ K around the bolometric luminosity peak. Analogous values of peak-light temperature have also been obtained from measurements of several rapidly evolving SLSNe~I \citep{2013ApJ...770..128I}. Based on the blackbody curve, such high temperatures correspond $g-r \sim -0.2$ mag and $B-V \sim -0.1$ mag.

In Figure~\ref{fig: comp_color}, most of the linear lines fitted to the $B-V$ and $g-r$ colour evolution for both the slowly and rapidly declining SLSNe~I (except for the $g - r$ colour of iPTF13ehe) tend to converge at the phase around the optical peak. \citet{2016ApJ...826...39N} apply the K-correction to the magnitudes and show a range of $-0.4\lesssim g-r\lesssim-0.1$ mag around peak. Although a larger sample of SLSNe~I appears to show a greater spread in peak-light colours, most of the values estimated by \citet{2018ApJ...860..100D} are roughly consistent with our results given their large uncertainties. Therefore, at least a portion of SLSNe~I likely have similar temperatures at maximum light. Such high temperatures at early phases are consistent with the absorption model constructed by \citet{2018ApJ...855....2Q}, which reproduces the strong O\,\textsc{ii} features by assuming a temperature of 15000 K. For those SLSNe~I with much lower temperatures, O\,\textsc{ii} features could be attributed to non-thermal excitation by a magnetar \citep{2016MNRAS.458.3455M}. At $t \approx 50$ d from the $r$-band maximum, the photosphere of SN 2018hti cooled below 9000~K, and eventually it remained at 7000--8500 K. We take the uncertainty of the temperature measurements into account and adopt $T \approx 7300$ K as the plateau temperature, at which the photosphere recedes inside the ejecta.

\subsection{Physical model}

There are many models proposed to explain the luminosity evolution of SLSNe. Popular ones include the magnetar-powered model (e.g., \citealp{2007ApJ...666.1069M, 2010ApJ...717..245K, 2010ApJ...719L.204W, 2016ApJ...821...22W}), the fallback accretion-powered model (e.g., \citealp{2013ApJ...772...30D, 2018ApJ...867..113M, 2018MNRAS.475L..11M}), and the CSM interaction-powered model (e.g., \citealp{2012ApJ...746..121C, 2013ApJ...773...76C, 2017ApJ...851L..14W, 2018ApJ...856...59L}). In the context of fallback accretion, unusually large amounts of accreted materials are needed for PTF12dam, iPTF13ehe, and SN 2015bn \citep{2018ApJ...867..113M}.

Given that SN 2018hti has a slowly declining light curve similar to these three SLSNe~I, fallback accretion-powered model might not be applied to this SN. SN 2018hti lacks narrow emission lines that indicate the interaction process. However, the possibility of CSM interaction cannot be completely ruled out. As an example, SN 2017egm is found to have a sharp peak in the bolometric light curve that is reminiscent of CSM interaction \citep{2017ApJ...851L..14W}, but no narrow emission features were detected in its spectra. Compared with the other two models, the magnetar-powered model is widely used to explain the photometric properties of SLSNe~I.

In this paper, we adopt a simplified version of the magnetar-powered model proposed by \citet{2017ApJ...850...55N} to fit the bolometric light curve of SN 2018hti. This model contains the following parameters: (1) explosion time ($T_0$), (2) mass ($M_\mathrm{NS}$), dipole magnetic field strength ($B$), and initial spin period ($P_0$) of a magnetar; (3) mass ($M_\mathrm{ej}$), velocity ($v_\mathrm{ej}$), and photospheric recession temperature ($T_\mathrm{phot}$) of the SN ejecta; and (4) opacities to optical photons ($\kappa$) and to gamma-ray photons from the magnetar wind ($\kappa_{\gamma,m}$). The mass of magnetar is assumed to be a typical value of $1.4$M$_\odot$, while the critical temperature is set\
 to be $T_\mathrm{phot}=7300$ K as suggested by the analysis in Section~\ref{SubSec: PhotEvo}. We fix $\kappa=0.2$ cm$^2$ g$^{-1}$ and $\kappa_{\gamma,m}=0.01$ cm$^2$ g$^{-1}$. The remaining parameters can be constrained by the evolution of $L_\mathrm{bol}$ and $T_\mathrm{eff}$.  We show the fitting results in Figure~\ref{fig: LT_fit} with $\chi^2/\mathrm{d.o.f}=0.6$, corresponding to $T_0=$ MJD 58405.1, $B=1.8\times10^{13}$~G, $P_0=1.8$~ms, $M_\mathrm{ej}=5.8~$M$_\odot$, and $v_\mathrm{ej}=6800$ km s$^{-1}$.

The fitting suggests that it takes $\sim50$ days for SN 2018hti to reach its peak luminosity, consistent with the typical rise time for SLSNe~I (see \citealp{2019NatAs...3..697I}, and references therein). With these parameters, we estimate the kinetic energy to be $E_\mathrm{k} \approx 2.6\times10^{51}$ erg for SN 2018hti. Most of our results are highly consistent with the median values from a larger sample, which are $P_0=2.4^{+1.6}_{-1.2}$~ms, $M_\mathrm{ej}=4.8^{+8.1}_{-2.6}~$M$_\odot$, and $E_\mathrm{k}=3.9^{+5.9}_{-2}\times10^{51}$ erg \citep{2017ApJ...850...55N}. The inferred magnetic field of SN 2018hti is close to that of the two well-studied slowly declining sample, PTF09cnd and PTF12dam, although all three objects have a relatively low magnetic field compared to the whole sample ((1--3) $\times 10^{13}$~G vs. $8^{+11}_{-6} \times 10^{13}$~G). Moreover, similar values of $P_0$, $M_\mathrm{ej}$, and $v_\mathrm{ej}$ are also derived for PTF09cnd and PTF12dam \citep{2017ApJ...850...55N}, suggesting similar explosion physics.

Combining the mass of the ejecta and that of the magnetar, one can derive the carbon-oxygen (CO) core mass of the progenitor to be $\sim 7.2~$M$_\odot$. In the case of single-star evolution \citep{2006A&A...460..199Y}, SN 2018hti could originate from a zero-age main sequence (ZAMS) star with an initial mass $25$--40 M$_\odot$. This is consistent with the mass range inferred for a typical Wolf-Rayet (WR) star \citep{2007ARA&A..45..177C}, which has a powerful outward stellar wind and provides a natural explanation for the absence of H in the SN spectra. The metallicity and the rotation of the progenitor star have significant influence on the stellar evolution. Based on the simulation in the case of low metallicity, $Z=0.3~Z_\odot$ (see Table 4 of \citealp{2006A&A...460..199Y}), the star related to a CO core of $\sim 7~M_{\odot}$ can have a mass of $\sim16$--25~M$_\odot$ and rotates at a fraction of the Keplerian velocity at the equator. At a similar low metallicity (i.e., $Z=0.4~Z_\odot$), a rotating B-type star of 15~M$_\odot$ was reported to hardly form a CO core over 4~M$_\odot$ \citep{2013A&A...553A..24G}. For a higher metallicity, $Z=Z_\odot$, an initial stellar mass of $\sim25$--32~M$_\odot$ possibly leads to a CO core of $\sim7~$M$_\odot$ \citep{2012A&A...542A..29G}. Thus, taking the rotation effect into account, the progenitor of SN 2018hti may have a lower initial mass, $\sim16$--25~M$_\odot$.

\citet{2015Natur.528..376M} propose that the magnetic field of rapidly spinning proto-neutron star can be significantly amplified due to the magnetorotational instability. This scenario is consistent with our fitting results that require a central magnetar with a spin period of around 1~ms. Since the spin of a proto-neutron star is widely believed to have a dependence on the initial rotation rate of the iron core before explosion \citep{2006ApJS..164..130O}, it is possible that SN 2018hti arose from the collapse of a rotating progenitor star. Alternatively, binary interaction (such as merger and mass transfer) is an efficient mechanism for the progenitor to gain its spin angular momentum \citep{2013ApJ...764..166D}. If it exploded in a binary system, SN 2018hti might originate from a star with $\gtrsim25~$M$_\odot$, which could result in the CO core mass of $>4~$M$_\odot$ based on the simulation of \citet{2010ApJ...725..940Y}. Lower-mass binary stars are proposed as candidate progenitors of SNe~Ic \citep{2016MNRAS.457..328L}, which show some post-peak spectral similarity to SN 2018hti, as discussed in Section~\ref{SubSec: SpecEvo}. However, the relationship between the progenitors of these two subtypes remains mysterious, since SN 2018hti has a higher peak luminosity, broader light curves, and a relatively lower-metallicity host.

\begin{figure}
\center
\includegraphics[angle=0,width=1\textwidth]{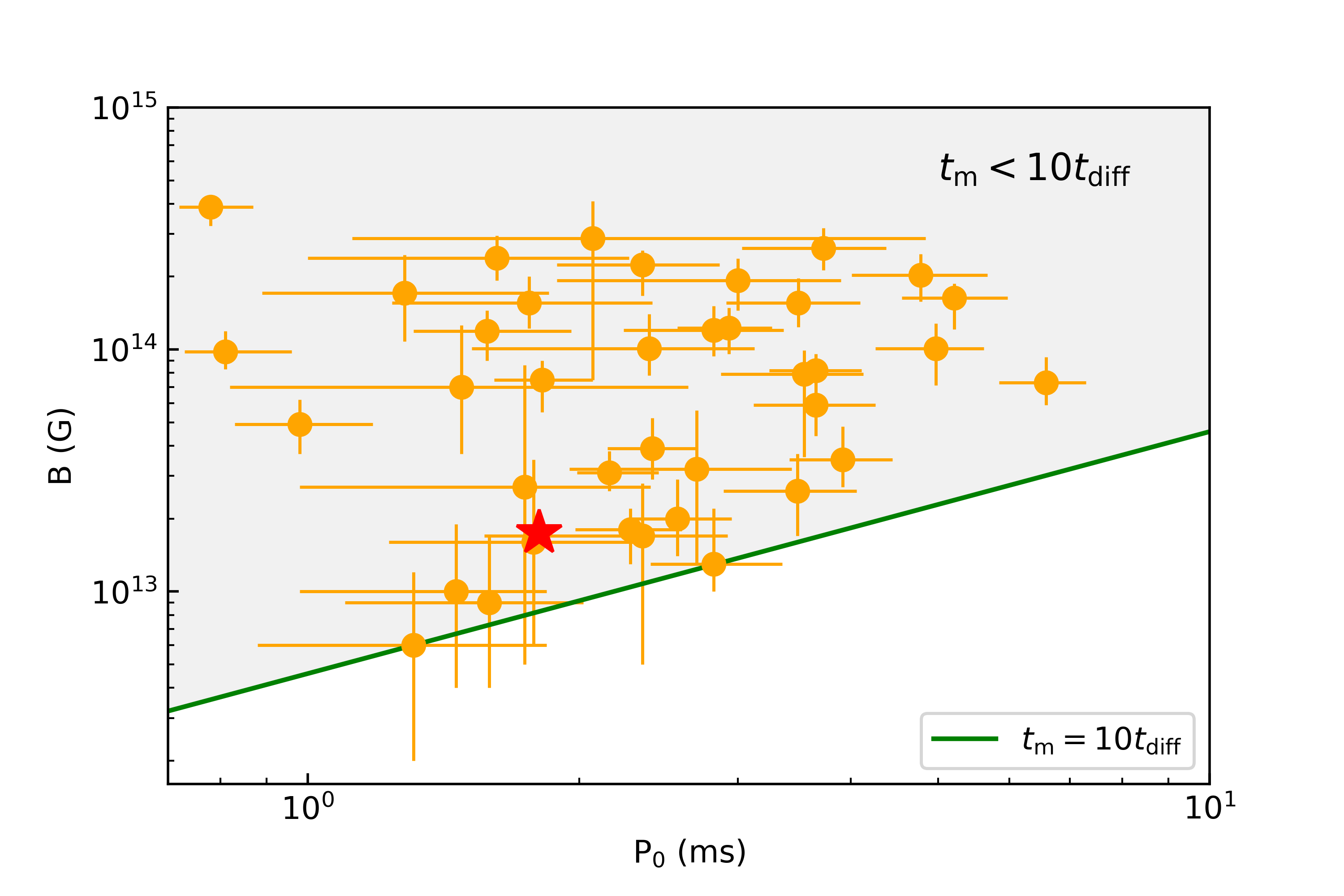}
\caption{Correlation between the initial spin period $P_0$ and magnetic field $B$ of the magnetars accounting for the emission of SN 2018hti (star) and the other SLSNe~I (circles; \citealp{2017ApJ...850...55N}). The solid line corresponds to the equation $t_\mathrm{m} = 10\,t_\mathrm{diff}$ (see text for details).}
\label{fig: PB}
\end{figure}

As shown in Figure~\ref{fig: PB}, magnetars that can power SLSNe~I tend to have spin periods of $P_0\approx0.7$--8~ms and magnetic fields of $B\approx6\times10^{12}$--$4\times10^{14}$~G, consistent with our results. Note that SLSNe~I are seldom found with $B\gtrsim 10^{15}$~G, as a very strong magnetic field is supposed to produce luminous events with very short timescales. In other words, an easy-to-detect SLSN~I requires sufficiently high injection power provided by a long-lived central engine. \citet{2017ApJ...850...55N} found that most of the $P_0 - B$ parameters of magnetars that accounts for the observed properties of SLSNe~I fall between the lines where the spin-down timescale ($t_\mathrm{m}$)\footnote{$t_\mathrm{m}=3c^3IP_0^2/4\pi^2B^2R_\mathrm{NS}^6$, where $c$ is the speed of light, $I$ is the moment of inertia, and $R_\mathrm{NS}$ is the radius of magnetar.} of a magnetar is proportional to the diffusion timescale ($t_\mathrm{diff}$)\footnote{$t_\mathrm{diff}=(2\kappa M_\mathrm{ej}/\beta c v\mathrm{ej})^{1/2}$ with $\beta=13.8$.}, or $0.1 \lesssim t_\mathrm{m}/t_\mathrm{diff} \lesssim 10$. Based on an Arnett-like semi-analytic model, the peak timescale ($t_\mathrm{p}$) scales with the diffusion timescale ($t_\mathrm{p}\propto t_\mathrm{diff}$), and the coefficient between these two timescales is usually $\sim0.2$--1 as $t_\mathrm{m}/t_\mathrm{diff}$ varies from $10^{-2}$ to $10^{2}$ \citep{2019ApJ...878...56K}. Thus, the $t_\mathrm{m}-t_\mathrm{diff}$ distribution corresponds to the relation between $t_\mathrm{m}$ and $t_\mathrm{p}$, which requires that the peak time of the light curve is comparable to the energy-injection timescale.

From the perspective of observations, the detectability of a SLSN increases when the SN ejecta get a large amount of energy input shortly before or during the diffusion phase. However, if the spin-down timescale is much shorter than the diffusion timescale, most of the injection energy would contribute to the adiabatic expansion of the ejecta at early phases instead of increasing the radiated luminosity. An event of such a scenario would be difficult to detect as an SLSN. SN 2018hti is located near the lower end of the distribution of both the $P_0$ and $B$ parameters, and hence it is close to the line of $t_\mathrm{m}=10\,t_\mathrm{diff}$. This indicates that the central magnetar has a relatively long spin-down timescale, which provides a natural explanation for the slowly declining behaviour seen in SN 2018hti. \citet{2017ApJ...840...12Y} proposed that a critical magnetic field of $\sim2\times10^{14}$~G could possibly divide SLSNe~I into slowly and rapidly evolving subgroups. Based on this criterion, the inferred field strength of $10^{13}$~G for SN 2018hti is also in agreement with that of the slowly evolving events.

\begin{figure}
\center
\includegraphics[angle=0,width=1\textwidth]{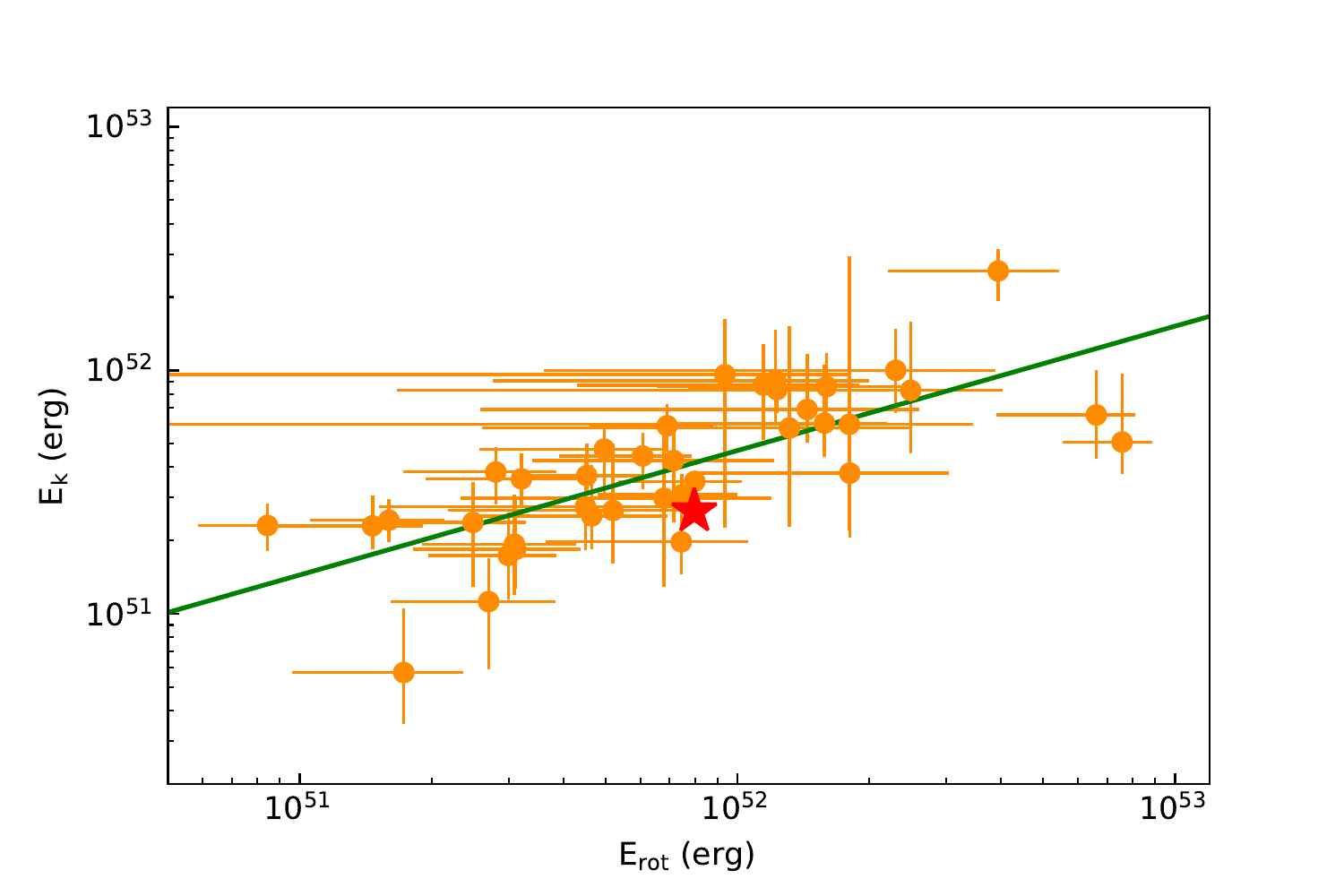}
\caption{Correlation between rotational energy $E_\mathrm{rot}$ and the kinetic energy $E_\mathrm{k}$ for SN 2018hti (stars) and other SLSNe~I (circles). The solid line represents the best fit to these two parameters, log\,$E_\mathrm{k} = 25.1 + 0.5$\,log\,$E_\mathrm{rot}$.}
\label{fig: ER_EK}
\end{figure}

For SN 2018hti, the initial rotation energy of the magnetar is calculated as $E_\mathrm{rot} \approx 8\times10^{51}$ erg. Figure~\ref{fig: ER_EK} shows the distribution of $E_\mathrm{rot}$ and $E_\mathrm{k}$ for SN 2018hti and the comparison sample. The energies of the comparison SLSNe~I are calculated using the model parameters given by \citet{2017ApJ...850...55N}. Regardless of large uncertainties for some sample, the kinematic energy $E_\mathrm{k}$ generally shows a positive correlation with the initial rotational energy $E_\mathrm{rot}$, as described by the empirical formula log\,$E_\mathrm{k} = 25.1 +  0.5$\,log\,$E_\mathrm{rot}$. This suggests that a nascent magnetar with a large spin energy is usually accompanied by a giant explosive ejection, which is consistent with the simulation result on the effect of rotation in the explosion of core-collapse SNe (\citealp{2005ApJ...620..861T}). The initial rotation energy of the progenitor star could provide an extra source to power the SN; also, a faster rotation will result in a lower effective gravity, which makes it easier to release matter from the central compact object. In the scenario of an alternative mechanism to drive an SN, such as a magnetorotational core-collapse SN (e.g., \citealp{1970ApJ...161..541L, 2007ApJ...664..416B, 2014ApJ...785L..29M}), we also expect that a rapidly spinning central object is associated with an intense ejection. Moreover, a fraction of the rotation energy deposited in magnetars can be efficiently extracted to accelerate ejecta \citep{2016ApJ...821...22W}.

\section{Conclusion}
\label{Sec: Concl}
SN 2018hti is a Type I superluminous supernova harboured in a faint dwarf galaxy with a metallicity of $\sim0.3\,Z_\odot$ and an SFR of $\sim0.3\,$M$_\odot$ yr$^{-1}$. We present extensive ultraviolet and optical photometric data obtained with the AZT, LJT, {\it Swift}, and TNT, spanning over a year after discovery. SN 2018hti has an absolute $g$-band magnitude of $-22.2$ at maximum brightness and a peak bolometric luminosity of $\sim3.5\times10^{44}$ erg s$^{-1}$. Both of these values are large enough to be above the photometric threshold of SLSNe~I. We also compare this SLSN~I to a well-observed sample, suggesting that SN 2018hti belongs to the slowly fading subclass like PTF12dam as evidenced by the slow light and colour curve evolution after the peak. The classification of SLSN~I is also favoured by the spectral evolution which exhibits strong C and O absorption features, especially O\,\textsc{ii} blends, superimposed on blue continua at early phases. SN 2018hti exhibited post-peak spectroscopic evolution analogous to that of SNe~Ic like SN 2002ap and SN 2007gr, but at a rate that is two times slower than the latter. This might be related to large ejecta mass and sustainable energy source for SN 2018hti and other SLSNe~I.

We model the bolometric luminosity and effective temperature evolution of SN 2018hti, and find that they are well consistent with that inferred from the SN ejecta of a mass of $5.8~$M$_\odot$ that is powered by a millisecond magnetar. Assuming a mass of $1.4\,$M$_\odot$ for the nascent magnetar, the pre-explosion CO core should have a mass of $7.2\,$M$_\odot$. The immediate progenitor of SN 2018hti might evolve from a ZAMS star with $\sim25$--40~M$_\odot$ or from a rotating star with an initial mass of $\sim16$--25~M$_\odot$. The progenitor could also be a star with $M_\mathrm{ZAMS}\gtrsim25~$M$_\odot$ in a binary system. In all of the above cases, the progenitor of SN 2018hti may not necessarily be much more massive than those of SNe~Ic except for lower metallicity and/or faster rotation velocities. The magnetar responsible for the emission of SN 2018hti is required to have an initial spin period of 1.8~ms and a modest magnetic field of $1.8\times10^{13}$~G, both of which are close to those of well-studied, slowly fading SLSNe~I PTF09cnd and PTF12dam. We estimate that the rotation energy of this magnetar to be $8\times10^{51}$ erg and the kinetic energy of the SN ejecta to be $2.6\times10^{51}$ erg. Based on a larger sample of SLSNe~I, we found that rapidly rotating magnetars tend to be surrounded by SN ejecta with large kinetic energies.

In conclusion, SN 2018hti exhibits photometric and spectroscopic properties that are distinct from those of normal stripped-envelope core-collapse SNe at early phases. This new sample contributes to a comprehensive understanding of the connections between the properties of massive stars and the resulting SNe. We note that several SLSNe~I such as LSQ14bdq and SN 2006oz reveal early bumps ahead of the primary peaks \citep{2012A&A...541A.129L, 2015ApJ...807L..18N, 2016MNRAS.457L..79N}. It is not clear whether this early bump exists in SN 2018hti because of the lack of early observations. The Zwicky Transient Facility (ZTF) survey with high cadence and large view of field \citep{2019PASP..131a8002B, 2019PASP..131g8001G} makes some progress in observations of infant explosions, which can provide clues to the nature of progenitor stars. ZTF has detected four SLSNe~I during commissioning and the early survey of two months, and observed at least 66~SLSNe since its full operation (e.g., \citealp{2019arXiv191002968L, 2019TNSAN..45....1Y}). Observations of a larger sample can be expected from ZTF, and will further develop statistical and theoretical studies regarding SLSNe~I. With the advent of the new era of multimessenger astronomy, we expect  significant progress in observations at other wavelengths (such as X-ray and radio), providing further insights into the physics of SLSNe~I and their possible connections with long-duration gamma-ray bursts and normal core-collapse SNe.

\section*{Acknowledgements}

We acknowledge the support of the staff of XLT, LJT, APO, AZT, and the Lick and Keck Observatories for assistance with the observations. Funding for the LJT has been provided by the Chinese Academy of Sciences and the People's Government of Yunnan Province. The LJT is jointly operated and administrated by Yunnan Observatories and Center for Astronomical Mega-Science, CAS. Some of the data presented herein were obtained at the W. M. Keck Observatory, which is operated as a scientific partnership among the California Institute of Technology, the University of California, and NASA; the observatory was made possible by the generous financial support of the W. M. Keck Foundation. A major upgrade of the Kast spectrograph on the Shane 3~m telescope at Lick Observatory was made possible through generous gifts from William and Marina Kast as well as the Heising-Simons Foundation. Research at Lick Observatory is partially supported by a generous gift from Google.

This work is supported by the National Natural Science Foundation of China (NSFC grants 11633002, 11761141001, and 11325313) and the National Program on Key Research and Development Project (grant 2016YFA0400803). J.J.Z. is supported by the NSFC (grants 11773067 and 11403096), the Key Research Program of the CAS (grant KJZD-EW-M06), the Youth Innovation Promotion Association of the CAS (grant 2018081), and the CAS ``Light of West China'' Program. T.M.Z. is supported by the NSFC (grant 11203034). This work was also partially supported by the Open Project Program of the Key Laboratory of Optical Astronomy, National Astronomical Observatories, Chinese Academy of Sciences. L.J.W. acknowledges support from the National Program on Key Research and Development Project of China (grant 2016YFA0400801). L.D.L. is supported by  the National Postdoctoral Program for Innovative Talents (grant BX20190044), China Postdoctoral Science Foundation (grant 2019M660515), and ``LiYun'' postdoctoral fellow of Beijing Normal University. Support for A.V.F.'s group has been provided by the TABASGO Foundation, the Christopher R. Redlich Fund, and the Miller Institute for Basic Research in Science (U.C. Berkeley).

\section*{Data availability}

The photometry of SN 2018hti is available in Tables 2-5, and the spectra will be shared on reasonable request to the corresponding authors.

\section*{Software}
IRAF \citep{Tody1986, Tody1993} and SExtractor \citep{1996A&AS..117..393B}.

{}

\clearpage

\begin{table}
\hspace{-1cm}
\caption{Magnitudes of reference stars in the field of SN 2018hti.}
\begin{tabular}{cccccccc}
\hline
Star & $\alpha$(J2000)& $\delta$(J2000) &$B$ & $V$ & $g$ & $r$ & $i$\\
\hline
1 & 3$^\mathrm{h}$40$^\mathrm{m}$35.72$^\mathrm{s}$ & 11\degr43\arcmin51.28\arcsec & 15.814(61) & 14.431(35) & 15.008(4) & --  & -- \\
2 & 3$^\mathrm{h}$40$^\mathrm{m}$36.39$^\mathrm{s}$ & 11\degr45\arcmin35.76\arcsec & 15.914(48) & 14.948(19) & 15.297(3) & 14.562(2) & 14.198(2) \\
3 & 3$^\mathrm{h}$40$^\mathrm{m}$44.63$^\mathrm{s}$ & 11\degr45\arcmin54.13\arcsec & 15.535(49) & 14.196(23) & 14.722(3) &  --  & -- \\
4 & 3$^\mathrm{h}$40$^\mathrm{m}$44.86$^\mathrm{s}$ & 11\degr41\arcmin27.94\arcsec & 17.085(181) & 16.326(113) & 16.778(4) & 16.199(4) & 15.899(4) \\
5 & 3$^\mathrm{h}$40$^\mathrm{m}$46.21$^\mathrm{s}$ & 11\degr46\arcmin28.36\arcsec & 16.139(93) & 14.933(20) & 15.373(2) & 14.472(3) & -- \\
6 & 3$^\mathrm{h}$40$^\mathrm{m}$53.93$^\mathrm{s}$ & 11\degr41\arcmin29.71\arcsec & 14.629(31) & 13.762(16) & 14.087(2) & -- & -- \\
7 & 3$^\mathrm{h}$40$^\mathrm{m}$55.71$^\mathrm{s}$ & 11\degr49\arcmin45.31\arcsec & 14.835(24) & 13.932(8) & 14.267(3)  & --& -- \\
8 & 3$^\mathrm{h}$41$^\mathrm{m}$2.73$^\mathrm{s}$ & 11\degr51\arcmin34.2\arcsec & 15.912(81) & 15.237(49) & 15.451(4) & 15.056(2)  & 14.883(4) \\
9 & 3$^\mathrm{h}$41$^\mathrm{m}$5.09$^\mathrm{s}$ & 11\degr45\arcmin34.66\arcsec & 16.15(100) & 15.094(59) & 15.475(2) & 14.670(2)  & 14.272(10) \\
10 & 3$^\mathrm{h}$41$^\mathrm{m}$9.75$^\mathrm{s}$ & 11\degr46\arcmin59.66\arcsec & 14.036(29) & 13.077(17) & -- & -- & -- \\
11 & 3$^\mathrm{h}$40$^\mathrm{m}$40.85$^\mathrm{s}$ & 11\degr42\arcmin30.6\arcsec & -- & -- & 17.338(6) & 16.357(18) & 15.937(17)  \\
12 & 3$^\mathrm{h}$41$^\mathrm{m}$17.99$^\mathrm{s}$ & 11\degr44\arcmin7.27\arcsec & -- & -- & 17.333(4) & 16.237(4) & 15.716(3) \\
13 & 3$^\mathrm{h}$40$^\mathrm{m}$43.3$^\mathrm{s}$ & 11\degr44\arcmin13.91\arcsec & -- & -- & -- & 17.458(4) & 16.794(5) \\
14 & 3$^\mathrm{h}$41$^\mathrm{m}$8.94$^\mathrm{s}$ & 11\degr42\arcmin39.19\arcsec & -- & -- & 17.146(2) &  16.422(2) & 16.074(2) \\
15 & 3$^\mathrm{h}$41$^\mathrm{m}$9.8$^\mathrm{s}$ & 11\degr43\arcmin8.65\arcsec & -- & -- & 17.229(5) & 16.315(4) & -- \\
16 & 3$^\mathrm{h}$41$^\mathrm{m}$14.19$^\mathrm{s}$ & 11\degr48\arcmin27.3\arcsec & -- & -- & -- & 16.784(2) & 16.421(3)  \\
\hline
\end{tabular}
\label{Table: referstar}
\\{Note: The numbers in the bracket are in unit of 0.001 mag. See Figure~\ref{fig: FinderChart} for the finder chart of SN 2018hti and most of the reference stars.}

\end{table}

\begin{table}
\hspace{-1cm}
\caption{SN 2018hti photometry from Lijiang 2.4-m telescope.}
\begin{tabular}{cccccc}
\hline
MJD & $B$ & $V$ & $g$ & $r$ & $i$\\
\hline
58429.78 & 17.94(3) & 17.52(2) & 17.64(3) & 17.57(1) & 17.71(2)\\ 
58431.7 & 17.75(3) & 17.32(2) & 17.41(2) & 17.36(1) & 17.36(1)\\ 
58434.73 & 17.48(3) & 17.07(2) & 17.16(1) & -- & 17.17(1)\\ 
58436.73 & 17.38(3) & 16.96(2) & 17.04(2) & 16.97(1) & 17.07(1)\\ 
58438.71 & 17.24(3) & 16.82(2) & 16.94(5) & 16.87(1) & 16.92(1)\\ 
58439.78 & 17.19(3) & 16.78(2) & 16.86(4) & 16.79(1) & 16.82(1)\\ 
58440.69 & 17.15(3) & 16.73(2) & 16.85(2) & 16.80(2) & 16.78(2)\\ 
58441.73 & 17.11(3) & 16.69(2) & 16.80(2) & 16.74(2) & 16.74(1)\\ 
58446.73 & 17.01(3) & 16.58(3) & -- & -- & --\\ 
58448.75 & 16.97(4) & 16.59(4) & -- & 16.52(1) & 16.48(1)\\ 
58449.75 & 16.91(3) & 16.51(4) & 16.56(4) & 16.51(2) & 16.49(1)\\ 
58451.72 & 16.89(3) & 16.48(2) & 16.53(4) & 16.44(1) & 16.44(1)\\ 
58456.72 & 16.80(3) & 16.40(2) & 16.45(5) & 16.39(1) & 16.35(1)\\ 
58457.69 & 16.80(3) & 16.40(2) & 16.45(4) & 16.39(1) & 16.34(1)\\ 
58460.71 & 16.80(3) & 16.41(2) & 16.45(2) & 16.38(1) & 16.39(1)\\ 
58492.61 & 17.12(3) & 16.61(2) & 16.75(3) & 16.56(1) & 16.51(1)\\ 
58497.64 & 17.19(3) & 16.64(4) & -- & -- & --\\ 
58516.54 & 17.72(3) & 17.02(4) & 17.28(2) & 16.94(1) & 16.83(1)\\ 
58528.5 & -- & 17.24(3) & -- & -- & --\\ 
58556.55 & 18.57(7) & 17.82(7) & -- & -- & --\\ 
58570.52 & 19.13(10) & 18.28(5) & 18.89(38) & 18.17(4) & 18.12(4)\\ 
58809.65 & -- & -- & 21.33(10) & 20.00(6) & 20.71(6)\\ 
\hline
\end{tabular}
\label{Table: photo_LJT}
\\{Note: The numbers in the bracket are in unit of 0.01 mag.}
\end{table}

\begin{table}
\caption{SN 2018hti photometry from Tsinghua-NAOC 0.8-m telescope.}
\begin{tabular}{cccccc|cccccc}
\hline
 MJD & $B$ & $V$ & $g$ & $r$ & $i$ &
 MJD & $B$ & $V$ & $g$ & $r$ & $i$  \\
\hline
58431.76 & 17.67(2) & 17.27(2) & 17.43(4) & 17.33(5) & 17.25(5) & 58489.61 & 17.04(2) & 16.56(2) & 16.72(4) & 16.51(4) & --\\ 
58438.52 & 17.17(4) & 16.78(2) & 16.95(3) & 16.82(6) & 16.80(5) & 58489.65 & -- & -- & -- & -- & 16.37(5)\\ 
58440.69 & 17.13(2) & 16.73(2) & 16.86(4) & 16.73(2) & 16.68(5) & 58490.66 & 17.04(3) & 16.54(2) & 16.74(5) & 16.50(6) & 16.34(8)\\ 
58441.61 & 16.95(7) & 16.69(4) & 16.87(4) & 16.71(6) & 16.66(8) & 58496.64 & 17.15(3) & 16.62(2) & 16.81(3) & 16.57(5) & 16.41(6)\\ 
58446.56 & 16.91(5) & 16.50(3) & 16.65(6) & 16.43(9) & 16.46(10) & 58498.58 & 17.28(6) & 16.70(3) & 16.99(5) & 16.63(9) & --\\ 
58447.6 & 16.91(4) & 16.56(2) & 16.63(5) & 16.53(9) & 16.42(5) & 58499.52 & 17.35(14) & -- & -- & -- & --\\ 
58449.58 & 16.88(2) & 16.47(2) & 16.56(4) & 16.49(4) & 16.39(5) & 58501.52 & 17.22(5) & 16.72(3) & 17.01(4) & 16.59(7) & 16.49(5)\\ 
58450.58 & 16.86(2) & 16.45(2) & 16.56(4) & 16.45(5) & 16.38(4) & 58503.51 & 17.17(9) & 16.74(4) & 17.03(5) & 16.70(13) & --\\ 
58451.58 & 16.85(2) & 16.43(2) & 16.56(4) & 16.42(4) & 16.36(3) & 58504.5 & 17.38(5) & 16.75(2) & 17.01(5) & 16.71(5) & 16.53(5)\\ 
58452.59 & 16.84(2) & 16.42(1) & 16.54(4) & 16.43(3) & 16.37(3) & 58511.49 & 17.54(2) & 16.89(1) & 17.11(4) & 16.80(3) & 16.65(3)\\ 
58455.71 & 16.80(2) & 16.38(2) & 16.50(5) & 16.42(6) & 16.30(7) & 58513.52 & 17.52(10) & 16.97(5) & 17.20(7) & 16.81(10) & 16.67(7)\\ 
58459.64 & 16.76(2) & 16.35(2) & 16.47(4) & 16.37(5) & 16.24(5) & 58514.52 & 17.64(3) & 16.90(2) & 17.19(5) & 16.88(4) & 16.64(5)\\ 
58461.63 & 16.78(2) & 16.37(2) & 16.50(4) & 16.35(4) & 16.26(4) & 58515.62 & 17.65(4) & 16.93(2) & 17.23(5) & 16.86(8) & 16.64(7)\\ 
58463.57 & 16.80(2) & 16.38(1) & 16.48(4) & 16.36(4) & 16.27(5) & 58525.48 & 17.91(6) & 17.11(3) & 17.51(6) & 17.00(5) & 16.82(9)\\ 
58464.54 & 16.79(2) & 16.37(2) & 16.47(4) & 16.34(6) & 16.26(5) & 58534.53 & 18.10(8) & 17.35(4) & -- & 17.36(6) & 17.03(5)\\ 
58465.57 & 16.81(2) & 16.40(1) & 16.51(4) & 16.37(4) & 16.27(3) & 58537.49 & 18.27(6) & 17.44(3) & 17.79(6) & 17.42(5) & --\\ 
58466.54 & 16.81(2) & 16.41(2) & 16.54(4) & 16.36(6) & 16.28(5) & 58539.49 & 18.24(7) & 17.46(4) & 17.85(3) & 17.44(7) & 17.12(6)\\ 
58467.68 & 16.84(2) & 16.39(2) & 16.51(5) & 16.39(5) & 16.27(5) & 58548.47 & 18.49(7) & 17.67(3) & 18.07(6) & 17.59(3) & 17.38(4)\\ 
58468.58 & 16.87(2) & 16.43(2) & 16.55(5) & 16.41(1) & 16.29(5) & 58567.47 & -- & -- & 18.80(11) & 17.67(13) & 17.79(8)\\ 
58469.64 & 16.90(3) & 16.43(2) & 16.58(7) & 16.41(5) & 16.27(6) & 58572.47 & -- & -- & -- & 17.73(9) & --\\ 
58476.6 & 16.96(5) & 16.46(2) & -- & 16.43(6) & 16.23(12) & 58728.85 & -- & -- & 20.47(7) & 19.32(6) & --\\ 
58477.58 & 16.95(3) & 16.49(2) & 16.57(6) & 16.44(7) & 16.32(7) & 58729.83 & -- & -- & -- & -- & 19.25(8)\\ 
58478.59 & 16.92(3) & 16.49(2) & 16.66(4) & 16.48(7) & 16.28(6) & 58732.79 & -- & -- & 20.73(6) & 19.52(5) & 19.28(7)\\ 
58479.56 & 16.96(2) & 16.45(2) & 16.61(4) & 16.44(5) & 16.30(6) & 58749.72 & -- & -- & -- & 19.72(10) & 19.33(10)\\ 
58480.58 & 16.94(2) & 16.48(2) & 16.65(4) & 16.42(5) & 16.28(7) & 58752.86 & -- & -- & -- & 19.77(8) & 19.50(9)\\ 
58481.58 & 16.95(2) & 16.50(2) & 16.66(5) & 16.47(4) & -- & 58756.84 & -- & -- & -- & -- & 19.39(12)\\ 
58482.58 & 16.95(2) & 16.51(1) & 16.64(3) & 16.47(3) & 16.32(4) & 58757.69 & -- & -- & -- & -- & 19.79(15)\\ 
58484.56 & 17.00(2) & 16.52(1) & 16.65(3) & 16.46(3) & 16.31(4) & 58758.71 & -- & -- & 21.11(12) & 19.50(7) & 19.26(8)\\ 
58485.57 & 17.03(2) & 16.54(1) & 16.67(4) & 16.50(3) & 16.35(5) & 58764.67 & -- & -- & -- & -- & 19.39(12)\\ 
58486.6 & 17.03(2) & 16.52(2) & 16.68(4) & 16.47(6) & 16.34(5) & 58782.83 & -- & -- & 21.05(14) & -- & --\\ 
58487.56 & 16.93(5) & 16.49(5) & 16.78(9) & -- & -- & 58787.76 & -- & -- & 21.32(13) & 19.69(8) & --\\ 
\hline
\end{tabular}
\label{Table: photo_TNT}
\\{Note: The numbers in the bracket are in unit of 0.01 mag.}
\end{table}

\begin{table}
\hspace{-1.0cm}
\caption{{\it Swift} photometry of SN 2018hti.}
\begin{tabular}{ccccccc}
\hline
MJD & $w2$ & $m2$ & $w1$ & $u$ & $b$ & $v$\\
\hline
58430.52 & 18.49(10) & 18.05(11) & 17.42(8) & 16.79(6) & 17.76(8) & 17.37(13)\\ 
58431.53 & 18.60(10) & 18.02(10) & 17.10(7) & 16.67(6) & 17.68(7) & 17.25(11)\\ 
58433.85 & 18.24(9) & 17.85(11) & 16.99(7) & 16.45(6) & 17.41(7) & 17.15(12)\\ 
58434.85 & 18.20(9) & 17.85(10) & 17.00(7) & 16.38(5) & 17.40(7) & 17.10(11)\\ 
58436.05 & 18.20(8) & 17.74(9) & 16.88(6) & 16.35(5) & 17.28(6) & 17.02(10)\\ 
58440.36 & 18.21(14) & 17.75(15) & 16.71(10) & 16.12(7) & 17.03(9) & 16.87(16)\\ 
58442.69 & 17.99(8) & 17.56(9) & 16.71(6) & 16.10(5) & 16.95(5) & 16.64(9)\\ 
58446.67 & 18.06(8) & 17.80(9) & 16.82(6) & 16.01(4) & 16.91(5) & 16.58(8)\\ 
58448.6 & 18.00(7) & 17.69(9) & 16.83(6) & 15.90(4) & 16.84(5) & 16.45(7)\\ 
58450.53 & 18.08(8) & 17.73(9) & 16.80(6) & 15.95(4) & 16.75(4) & 16.39(7)\\ 
58453.12 & 18.21(8) & 17.95(12) & 16.80(6) & 15.85(4) & 16.71(4) & 16.32(7)\\ 
58454.25 & 18.24(8) & 17.77(10) & 16.76(6) & 15.82(4) & 16.75(4) & 16.31(6)\\ 
58456.51 & 18.15(8) & 17.85(10) & 16.79(6) & 15.92(4) & 16.71(4) & 16.31(6)\\ 
58459.74 & 18.32(9) & 17.99(8) & 17.02(9) & 15.84(4) & 16.74(5) & 16.33(7)\\ 
58460.02 & 18.21(9) & 18.02(8) & 16.92(9) & 15.88(5) & 16.73(5) & 16.31(7)\\ 
58465.01 & 18.36(9) & 18.19(10) & 17.21(9) & 15.91(5) & 16.62(5) & 16.29(7)\\ 
58467.45 & 18.48(10) & 18.19(10) & 17.21(9) & 15.91(5) & 16.62(5) & 16.29(7)\\ 
58474.02 & 18.53(10) & 18.48(9) & 17.13(10) & 16.13(5) & 16.92(5) & 16.32(7)\\ 
58476.75 & 18.94(16) & 18.57(17) & 17.40(10) & 16.16(6) & 16.88(7) & 16.44(10)\\ 
58482.52 & 18.97(16) & 18.66(16) & 17.52(12) & 16.19(6) & 16.79(6) & 16.48(9)\\ 
58485.25 & 18.94(16) & 18.87(14) & 17.75(12) & 16.31(6) & 16.86(6) & 16.48(9)\\ 
58491.77 & 19.29(15) & 19.06(12) & 17.92(11) & 16.48(6) & 17.00(5) & 16.56(8)\\ 
58496.35 & 19.42(17) & 19.61(24) & 17.78(12) & 16.62(6) & 17.16(6) & 16.54(9)\\ 
58504.78 & 19.69(37) & 19.89(43) & 18.27(13) & 16.90(13) & 17.36(12) & 16.76(17)\\ 
58508.44 & 19.77(20) & 20.03(23) & 18.62(25) & 17.13(7) & 17.40(6) & 16.61(8)\\ 
58512.4 & 19.78(20) & 20.40(30) & 18.95(16) & 17.22(8) & 17.52(7) & 16.74(8)\\ 
58516.26 & 20.32(32) & 20.98(49) & 19.02(17) & 17.34(9) & 17.57(8) & 16.81(9)\\ 
58559.69 & >20.9 & >21.0 & 20.40(45) & >19.4 & 18.84(24) & 17.76(21)\\ 
58569.78 & >20.4 & >20.7 & >20.1 & >18.9 & 18.73(32) & 17.60(28)\\ 
\hline
\end{tabular}
\label{Table: photo_Swift}
\\{Note: The numbers in the bracket are in unit of 0.01 mag.}
\end{table}

\begin{table}
\caption{AZT photometry of SN 2018hti.}
\begin{tabular}{ccccc}
\hline
MJD & $B$ & $V$ & $R$ & $I$ \\
\hline
58715.98 & 20.85(13) & 19.63(8) & 19.45(7)   & 18.67(5)\\
\hline
\end{tabular}
\label{Table: photo_AZT}
\\{Note: Template subtraction was not performed for the AZT photometry. The numbers in the bracket are in unit of 0.01 mag.}
\end{table}

\begin{table}
\caption{Journal of spectroscopic observations of SN 2018hti.}
\begin{tabular}{ccccccc}
\hline
UT & MJD & Phase$^{a}$ & Telescope  & Instrument & Exp. time (s) & Range (\AA) \\
\hline
 2018-11-07 &  58429.7 & $-$34.8 & LJT & YFOSC(G3) &1953  & 3502$-$8769 \\
 2018-11-08 &  58430.6 & $-$33.9 & LJT & YFOSC(G3) & 2700 & 3506$-$8769 \\
 2018-11-09 &  58431.4 & $-$33.1 & Lick 3\,m & Kast & 3000 & 3614$-$10738 \\
 2018-11-12 &  58434.7 & $-$29.8 & LJT & YFOSC(G3)  & 2200 & 3501$-$8769 \\
 2018-11-17 &  58439.7 & $-$24.8 & LJT & YFOSC(G3)  & 2000 & 3502$-$8768 \\
 2018-11-18 &  58440.6 & $-$23.9 & XLT & BFOSC(G4)  & 3300 & 3855$-$8696 \\
 2018-11-19 &  58441.6 & $-$22.9 & XLT & BFOSC(G4)  & 3600 & 3848$-$8698 \\
 2018-11-29 &  58451.7 & $-$12.8 & LJT & YFOSC(G3)  & 1800 & 3503$-$8767 \\
 2018-12-05 &  58457.6 & $-$6.9 & LJT & YFOSC(G3)  & 2000 & 3497$-$8768 \\
 2018-12-09 &  58460.5 & $-$4.0 & XLT & BFOSC(G4)  & 3300 & 3966$-$8691 \\
 2018-12-23 &  58475.6 & $+$11.0 & LJT & YFOSC(G3)  & 2200  & 3506$-$8765 \\
 2019-01-04 &  58487.3 & $+$22.6 & Lick 3\,m & Kast & 1800 & 3614$-$10648 \\
 2019-01-06 &  58489.6 & $+$25.0 & XLT & BFOSC(G4)  & 3600 & 3951$-$8692 \\
 2019-01-14 &  58496.5 & $+$31.9 & LJT & YFOSC(G3)  & 2200 & 3496$-$8765 \\
 2019-01-27 &  58510.1 & $+$45.5 & Lick 3\,m & Kast & 2400 & 3614$-$10500 \\
 2019-02-02 &  58516.5 & $+$51.9 & LJT & YFOSC(G3)  & 2400 & 3504$-$8766 \\
 2019-02-08 &  58522.4 & $+$57.8 & XLT & BFOSC(G4)  & 3600 & 4366$-$8705 \\
 2019-02-24 &  58538.4 & $+$73.8 & XLT & BFOSC(G4)  & 3300 & 4372$-$8705 \\
 2019-03-14 &  58556.5 & $+$91.9 & LJT & YFOSC(G3)  & 2400 & 3500$-$8764 \\
 2019-04-04 & 58577.2 & $+$112.6 & Keck I & LRIS & 600 & 3500$-$10282 \\
 2019-08-25 & 58720.4 & $+$255.8 & APO 3.5-m & DIS & 1800 & 5747$-$9200\\
 2019-10-28 & 58784.6 & $+$319.9 & Keck I & LRIS & 615 & 3500$-$10287 \\
 \hline
\end{tabular}
\\$^{a}${Days with respect to the epoch of $r$-band maximum (MJD 58464.6).}
\label{Table: spec_Journal}
\end{table}

\begin{table}
\caption{Emission-line properties of the SN 2018hti host galaxy.}
\begin{tabular}{cccc}
\hline
Line & Flux ($10^{-17}$ erg cm$^{-2}$ s$^{-1}$) & FWHM (\AA)  & EW (\AA) \\
\hline
H$\alpha$ & 421.1$\pm$7.5 & 6.2$\pm$0.1 & 101.2$\pm$1.8 \\
H$\beta$ & 139.4$\pm$5.1 & 3.4$\pm$0.1 & 21.4$\pm$0.8 \\
H$\gamma$ & 56.2$\pm$6.0 & 3.4$\pm$0.3 & 6.4$\pm$0.7 \\
H$\delta$ & 24.2$\pm$6.5 & 2.7$\pm$0.6 & 2.6$\pm$0.7 \\
\text{[N~II]} $\lambda$6584 & 13.6$\pm$7.8 & 6.4$\pm$2.8 & 3.3$\pm$1.9 \\
\text{[O~II]} $\lambda$3727 & 271.0$\pm$12.7 & 4.9$\pm$0.2 & 32.2$\pm$1.5 \\
\text{[O~III]} $\lambda$5007 & 785.5$\pm$11.0 & 3.4$\pm$0.0 & 96.8$\pm$1.4 \\
\text{[O~III]} $\lambda$4959 & 260.8$\pm$5.4 & 3.5$\pm$0.1 & 35.9$\pm$0.7 \\
\text{[Ne~III]} $\lambda$3869 & 64.9$\pm$10.4 & 3.6$\pm$0.4 & 7.1$\pm$1.1 \\
\text{[S~II]} $\lambda$6717 & 22.6$\pm$2.8 & 6.5$\pm$0.6 & 6.6$\pm$0.8 \\
\text{[S~II]} $\lambda$6731 & 15.7$\pm$2.7 & 6.0$\pm$0.8 & 4.7$\pm$0.8 \\
\text{[S~III]} $\lambda$9069 & 21.5$\pm$2.1 & 5.4$\pm$0.4 & 9.5$\pm$0.9 \\
\hline
\end{tabular}
\label{Table: host_lines_fit}
\end{table}

\clearpage

\bsp	
\label{lastpage}
\end{document}